\documentclass{aa}

\usepackage{newtxtext,newtxmath}

\usepackage[T1]{fontenc}
\usepackage{ae,aecompl}

\usepackage{natbib}
\bibpunct{(}{)}{;}{a}{}{,} 

\usepackage{graphicx}	
\usepackage{amsmath}	
\usepackage{amssymb}	
\usepackage{xcolor}
\usepackage[normalem]{ulem}

\newcommand{\gal} {{\mathrm{gal}}}

\newcommand{\ngc} {\emph{nGC3}}
\newcommand{\gc} {\emph{GC3}}
\newcommand{\mgc} {\emph{mGC3}}

\newcommand{\mas} {mas\,yr$^{-1}$}
\newcommand{\kms} {km\,s$^{-1}$}
\newcommand{\Gaia} {\emph{Gaia} }
\newcommand{\bp}{G_{BP}}
\newcommand{\rp}{G_{RP}}
\def\spsi{{\sin(\psi)}}
\def\cpsi{{\cos(\psi)}}
\def\dpsidR{{{\partial\psi}\over{\partial R}}}
\def\dpsidth{{{\partial\psi}\over{\partial \theta}}}

\def\dpsidxl{({{\partial\psi}/{\partial x}})}
\def\dpsidyl{({{\partial\psi}/{\partial y}})}

\begin{document}
\title{\Gaia~kinematics reveal a complex lopsided and twisted Galactic disc warp}

\author{M. Romero-G\'omez\inst{\ref{inst1}} \and C. Mateu\inst{\ref{inst2},\ref{inst3}} \and L. Aguilar\inst{\ref{inst4}} \and F. Figueras\inst{\ref{inst1}} \and A. Castro-Ginard \inst{\ref{inst1}}}

\institute{
Institut de Ci\`encies del Cosmos, Universitat de Barcelona (IEEC-UB), Mart\'{i} i Franqu\`es 1, E-08028 Barcelona, Spain \email{merce.romero@ub.edu} \label{inst1}
\and
Departamento de Astronom\'{i}a, Instituto de F\'{i}sica, Universidad de la Rep\'{u}blica, Igu\'a 4225, CP 11400 Montevideo, Uruguay \label{inst2}
\and
Centro de Investigaciones de Astronom\'ia (CIDA), AP 264, M\'erida 5101-A, Venezuela
\label{inst3}
\and
Instituto de Astronom\'{i}a, Universidad Nacional Auton\'{o}ma de M\'{e}xico, Apdo. Postal 877, Ensenada, 22800 Baja California, Mexico \label{inst4}
}

\date{Accepted XXX. Received YYY; in original form ZZZ}

\abstract
{There are few warp kinematic models of the Galaxy able to characterise both structure and kinematics, since they require high accuracy at large distances. These models are necessary to shed some light to the lopsidedness of the warp and the twisting of the line-of-nodes of the stellar warp, already seen in gas and dust.
}
{We use the vertical information coming from the \Gaia~Data Release 2 astrometric data up to $G=20$~mag to characterise the structure of the Galactic warp, the related vertical motions and the dependency of the Galactic warp on the age.}
{We analyse two populations up to galactocentric distances of $16$~kpc, a young bright sample mainly formed by OB stars and an older one of Red Giant Branch (RGB) stars. We use two methods (the Pole Count Maps of Great Circle bands and galactic longitude - proper motion in latitude lines) based on the \Gaia~ observables, together with two dimensional projections of the positions and proper motions in the Galactic plane.}
{This work confirms the age dependency of the Galactic warp, both in positions and kinematics, being the height of the Galactic warp of the order of $0.2$~kpc for the OB sample and of $1.0$~kpc for the RGB at a galactocentric distance of $14$~kpc. Both methods find that the onset radius of the warp is $12\sim 13$~kpc for the OB sample and $10\sim 11$~kpc for the RGB. From the RGB sample, we find from galactocentric distances larger than $10$~kpc the line-of-nodes twists away from the Sun-anticentre line towards galactic azimuths $\sim 180-200\degr$ increasing with radius, though possibly influenced by extinction. Also, the RGB sample reveals a slightly lopsided stellar warp with $\sim 250$~pc difference between the up and down sides. The line of maximum of proper motions in latitude is systematically offset from the line-of-nodes estimated from the spatial data, which our warp models predict as a kinematic signature of lopsidedness. We also show a prominent wave-like pattern of a bending mode different in the OB and RGB samples. Both positions and kinematics also reveal substructures that might not be related to the large scale Galactic warp nor to the bending mode.}
{\Gaia~Data Release 2 data reveals a high degree of complexity in terms of both positions and velocities that triggers the need for complex kinematic models, flexible enough to combine both wave-like patterns and an S-shaped lopsided warp.  }


\maketitle

\section{Introduction}

Warped discs still represent a theoretical challenge nowadays. From surveys of edge-on galaxies, it is clear that $\sim 50-70 \%$ of spiral disc galaxies present stellar warped discs \citep[e.g.][]{Sanchez1990}, suggesting that they are long-lived, or repeatedly generated. Our Galaxy also presents a warped disc, first detected with 21-cm observations of the HI gas \citep[][among others]{Burke1957,Westerhout1957,Oort1958,Levine2006}, later in dust and stars using the Two Micron All Sky Survey infrared data \citep[][among others]{Freudenreich1994,Drimmel2001,LopezCorredoira2002,Reyle2009,Amores17}, and more recently using Cepheids \citep{Skowron2018,Chen2019}. Most of these studies dedicate their efforts to determine the morphology of the warped disc, that is, the galactocentric radius at which the disc starts bending, the phase angle of the line-of-nodes, its maximum amplitude and its possible dependence upon the tracer.

Very few studies have focused on analysing the warp kinematically, in the sense of finding the effect of a warped disc on the kinematics of the stars. The first kinematic analyses were conducted using \emph{Hipparcos} proper motions. \citet{Dehnen1998} selected a set of kinematically unbiased main sequence and giants stars and found evidence of the stellar warp when plotting the vertical velocity as a function of the tangential velocity. \citet{Drimmel2000} plotted the vertical velocity of the stars as a function of the galactocentric radius for a sample of OB stars, concluding that the kinematics of the stars towards the outer disc was inconsistent with the expectations from a long-lived non-precessing warp. Some years later, studies by \citet{Seabroke2007} as well as \citet{Bobylev2010,Bobylev2013} concluded the (then) available proper motion surveys do not allow complete studies of the Galactic warp. Using PPMXL proper motions and selecting disc Red Clump stars using 2MASS photometry, \citet{LopezCorredoira2014} found kinematic evidence of the stellar warp, by plotting the vertical velocity as a function of the galactocentric azimuth, concluding that their results cannot be reproduced by a population in statistical equilibrium. The quality of the \Gaia~ Data Release 1 and the Tycho-Gaia Astrometric Solution \citep[TGAS,][]{Michalik2015,Brown2016} proper motions allowed \citet{Schonrich2018} to review the study started by \citet{Dehnen1998}. Following a similar strategy, these authors find evidence of a kinematic signature of the Galactic stellar warp in the TGAS data in a cone towards the centre and anti-centre directions and estimate the onset of the warp at a guiding radius inside the Solar circle, $R_g\lesssim7$~kpc, in agreement with the previous work by \citet{Drimmel2001}. \citet{Carrillo2018} and \citet{Schonrich2018} also point out that the complexity of the data cannot be explained by a simple warp pattern, but also wave-like patterns, bending and breathing modes are reflected in the kinematic structure. The advent of the \Gaia~Data Release~2 (hereafter, GaiaDR2), and the availability of proper motions for $1.3$ billion sources  \citep{Brown2018}, has allowed to expand the study to larger volumes and fainter sources. \citet{Katz2018} has already shown the complexity of the vertical motion and \citet{Poggio2018} review the data selection of OB-type and Giants stars performed in \citet{Katz2018} and show the kinematic evidence of the Galactic warp in vertical velocity using two-dimensional maps.

The main challenge in our work is going one step forward. We want to use the spatial and kinematic data, together with models, to constrain the morphology of the Galactic warp. 
We assume this challenge by defining three Galactic warp models and using two characterisation methods developed by the authors specifically for this purpose. The complexity of the data leads to consider not only simple symmetric tilted rings models, but more complex asymmetric models, such as lopsided tilted rings or lopsided S-shaped warp models. We also use two methods to infer the structural parameters of the warp given the observables (positions and proper motions). The first is the LonKin method \citep{Abedi2015}, which provides information on the level of asymmetry of the warp and shape, when applied to different warp models. The second is \ngc, a method from the Great Circle Cell Counts (GC3) family \citep{Johnston1996,Mateu2011}, which searches for over-densities of stars along great circle cells. The latter has already been used in simulations by \citet{Abedietal14} to assess the \Gaia~ capabilities to derive the tilt angle, or amplitude of the warp, and the twist angle of the line-of-nodes. In this work, we apply both methods to two different populations from GaiaDR2, namely OB-stars and Red Giant Branch stars (similarly but not equally selected as in \citet{Poggio2018}). We define a strategy that allows us to select stars in both samples up to magnitude $G = 20$~mag in order to have enough statistics up to $R\le 16$~kpc and to disentangle the kinematic signature of the warp. Two tracer populations are necessary to study the dependency of morphology and kinematics on the age of the tracer, as recently suggested by \citet{Amores17}. We study the characteristics of the Galactic warp in terms of spatial and kinematic data of the two tracer populations, with intrinsically different ages, in order to give a first step towards understanding the origin of the Galactic warp. Additionally, the application of the methods to samples of test particles, evolved with the different models specified above, allows us to give an interpretation of the results in terms of shape, onset radius and tilt angle. 

The paper is organised as follows. In Sect.~\ref{sec:data_treatment} we give a brief description of the characterisation methods (Sect.~\ref{sec:methods}) used, we discuss the need to use an adequate distance estimator and we give the final choice (Sect.~\ref{sec:distance_estimators}),  and we describe the Gaia mock catalogues used to compare with the data (Sect.~\ref{sec:mocks}). In Sect.~\ref{sec:sample_selection}, we describe the selection of the working samples from GaiaDR2 data and give an initial characterisation. In Sect.~\ref{sec:spatial}, we study the spatial density distribution of both samples (Sect.~\ref{sec:spatial_maps}) and we give the first spatial characteristics of the Galactic warp (Sect.~\ref{sec:spatial_warp}). We then continue by studying the kinematic signature of the warp in Sect.~\ref{sec:kine}. We first focus on the kinematics in 2D projection maps (Sect.~\ref{sec:kine_maps}) and in Sects.~\ref{sec:lonkin_data} and \ref{sec:ngc_data} we apply the LonKin and \ngc~methods to the two populations extracted from the GaiaDR2 data, respectively. In Sect.~\ref{sec:discussion} we combine the spatial and kinematic distributions and we compare it with the models and data from the literature. Finally, in Sect.~\ref{sec:conclusion} we present our conclusions. This work is complemented with four appendices: in appendix~\ref{app:sample} we give further details on the sample selection; in appendix~\ref{app:models} we provide a full description of the warp models used; in appendix~\ref{app:veltrans} we detail the position and velocity transformations applied to a flat disc in order to warp it according to each of the three models; in appendix~\ref{app:predictions} we discuss the information the two methods provide when we apply them to synthetic data consisting of three different sets of particles, simulated with increasing complexity and reality.

\section{Methods and data treatment} 
\label{sec:data_treatment} 
In order to analyse the warp signature in the disc kinematics, we use methods previously developed by the authors \citep{Abedietal14}, namely the LonKin method and the Great Circle Cell Counts (hereafter, GC3) method and its variations. Both methods are designed to provide structural and kinematic information of the warp from the use of positions, distances and proper motions. First, in Sect.~\ref{sec:methods}, the methods are described and specify the requirements needed to apply them. Second, in Sect.~\ref{sec:distance_estimators}, we use mock catalogues to study and define the appropriate choice for the distance estimator used throughout this paper.

\subsection{Methods for warp detection and characterisation}
\label{sec:methods}
The LonKin method looks for the signature of a possible warped disc in a plot of proper motion in galactic latitude, $\mu_b$, as a function of the heliocentric galactic longitude $l$. The method as shown in this paper was developed in \citet{Abedi2015}, and its main advantage is that it uses directly the \Gaia~observables. The method is a variation of those used in \citet{Drimmel2000} and \citet{LopezCorredoira2014}, which plot the vertical velocity, W, as a function of the galactocentric radius and of the galactocentric azimuth, respectively. Therefore, the main advantage of LonKin with respect to these is that it works in the space of the observables, using proper motions instead of the vertical velocity, W. In the vertical axis we plot the proper motion in latitude corrected by the reflection of the solar motion, that is, with respect the Local Standard of Rest, $\mu_{b,LSR}$. In the horizontal axis, we plot the heliocentric galactic longitude, $l$, in segments of $20^\circ$. For each segment, we compute the median of the $\mu_{b,LSR}$ values and the lower ($15.85$-percentile) and upper ($84.15$-percentile) $1\sigma$ uncertainties on the estimation of the median as defined in the appendix A of \citet{Katz2018}, to which we add in quadrature the uncertainty given by the solar velocity. Since the Galactic warp is a galactocentric feature, we require to split the sample  into cylindrical galactocentric radial bins, for which we need to assume a Galactic solar radius and propagate the errors to the galactocentric frame to obtain the cylindrical galactocentric radius $R$. If the galactic disc is flat, the median $\mu_{b,LSR}$ should be constant and equal to zero, but if it is warped, a particular variation will be introduced as a function of $l$. If the disc is symmetrically warped and the line-of-nodes is aligned with the Sun-Galactic Centre line, the LonKin method predicts a maximum in $\mu_{b,LSR}$ in the anti-centre direction. If, on the other hand, the disc is asymmetric, this is, it is Lopsided, and the line-of-nodes, still on the Sun-Galactic Centre line, the maximum in $\mu_{b,LSR}$ is no longer in the anti-centre direction, but shifted towards longitudes coinciding the maximum warp amplitude (see detailed discussion in appendix~\ref{app:predictions}).

The family of Great Circle Cell Counts (hereafter, \gc) methods \citep{Johnston1996,Mateu2011} comprises different ways of searching for overdensities in great circle cells in the sky. In the most general version (\mgc) including full kinematic information --introduced in \citet{Mateu2011}-- the method sweeps over the sky counting how many stars have position and velocities lying in a great circle within a given tolerance, each great circle being defined uniquely by its normal vector or \emph{pole}. The all-sky sweep over all possible great circle cells results in a Pole Count Map (hereafter, PCM), a plot of the number of stars associated to each possible pole, and thus great circle cell, in the celestial sphere. To apply this method for the particular case of characterising the warp, PCMs are made for different galactocentric radial bins and the star counts are made in great circles defined in a Galactocentric reference frame. 
If the Galactic disc is flat, the peak in stellar density should be located in the North Galactic Pole of the PCM. If the disc is not flat, the peak of over-density should move in the PCM providing information on the tilt angle of the warp, as well as the azimuth (twist) of the line-of-nodes as a function of radii \citep[see][for detailed examples]{Abedietal14}. If the warp is lopsided, the signature in the PCM is not a single peak, but it has a shape that depends on the warp model (see detailed discussion in appendix~\ref{app:predictions}). In this work in particular, we will use the \ngc~method of the family, which uses 3D position information and proper motions, without requiring line-of-sight velocities. As we have shown in \citet{Abedietal14}, the use of the full velocity information in \mgc~ severely limits the sample spatial coverage and does not produce a significant improvement in the results.

\subsection{Selection of the distance estimator}
\label{sec:distance_estimators}

The Lonkin and \ngc~methods both start off by binning the sample in galactocentric distance. This is the most error prone step in both methods \citep[see][]{Abedietal14}, so, in order to reach as far as possible in the disc with the smallest possible distance bias it is crucial to have a well-behaved distance estimator. Thus, our analysis here is focused in estimating the galactocentric distance bias for stars binned in consecutive galactocentric rings, for different distance indicators under DR2 parallax error prescriptions  \citep{Lindegren2018}. This will also allow us to estimate the maximum galactocentric radius we can reach without introducing a significant bias.

We use the warped disc test particle simulation for RC stars up to $G=20$~mag from \citet{Abedietal14} and explore as distance indicators: the distance computed as the reciprocal of the parallax (excluding stars with negative parallaxes), the Bayesian indicator with an exponentially decreasing space density (EDSD) prior proposed by \citet{BailerJones15} and \citet{Astraatmadja16}, for  values of the scale length parameter $L=$1.35, 2.0 and 2.5~kpc; and the EDSD prior with a variable scale length $L(l,b)$ dependent upon the line of sight \citep[][hereafter BJ18]{BailerJones18}. We use the posterior mode as a point estimator for the distance from here on, as in BJ18. 
  
\begin{figure}
\begin{center}
 \includegraphics[width=0.95\columnwidth]{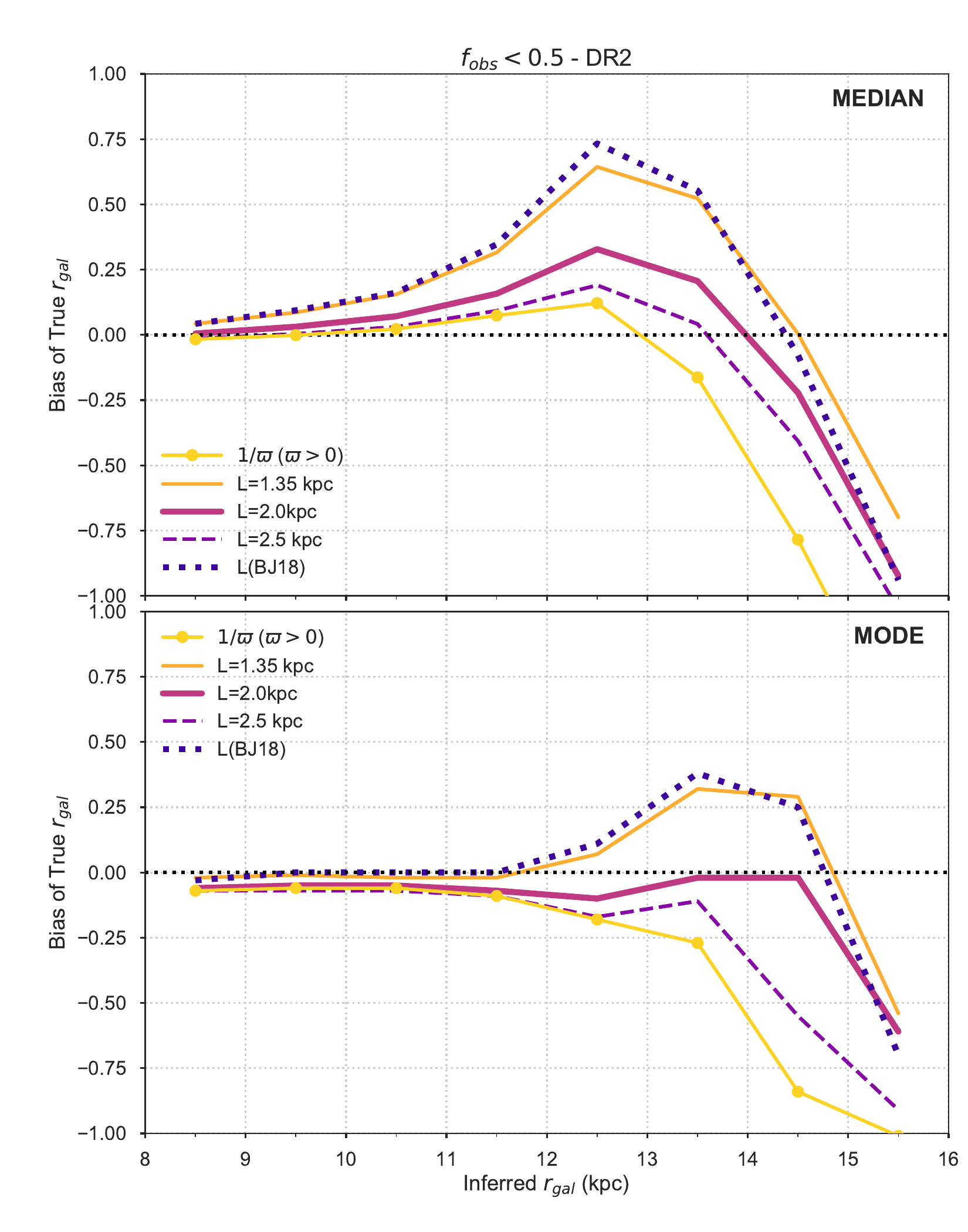} 
 \caption{Median (\emph{top}) and mode (\emph{bottom}) bias of the true $r_\gal$ distances of stars selected in bins of inferred $r_\gal$, for \Gaia~DR2 errors. Results for the reciprocal of the parallax are shown with the (yellow) solid+circle line; the prior scale lengths used with the Bayesian estimators are indicated with different colours as shown in the figure legends.}
\label{f:RtruDist_L_medmod_DR2}
\end{center}
\end{figure}

Figure~\ref{f:RtruDist_L_medmod_DR2} shows two summary statistics (median and mode) for
the bias in the inferred \emph{galactocentric} radius $r_\gal$, i.e. for stars in a given inferred $r_\gal$ bin, how deviated is the median or mode of their true $r_\gal$ distributions. In this plot we show the results obtained using stars with \emph{observed} parallax error smaller than 50\%, i.e. $f_{obs}=|\Delta\varpi/\varpi|<0.5$. We have checked that this $f_{obs}$ threshold allows us to exclude stars with arbitrarily large fractional parallax errors (both positive and negative), while keeping the galactocentric distance bias reasonably low. More details are given in Appendix~\ref{sec:app_dist}. We are aware that a change in the selected $f_{obs}$ threshold would affect the distribution of true distances of the stars selected in the sample and, therefore, the optimal prior scale length.

The figure shows the behaviour of the distributions is fairly stable up to $r_\gal\sim12-13$~kpc, beyond which there is an abrupt change.  For all estimators, the (median and mode) bias first increases up to that distance and then plummets to large negative values systematically. For any given estimator, it also shows the mode of the true distance distribution in a given bin is less biased and more stable than the median, at all distances.

For the \emph{median} $r_{\gal}$, the best performance up to the threshold radius ($\sim13$~kpc for RC stars) is obtained with the reciprocal of the parallax  and the Bayesian EDSD prior with $L=2.5$~kpc. At larger distances the $L=2$~kpc estimator performs slightly better than $L=2.5$~kpc and much better than the reciprocal of the parallax. In general, the BJ18 $L(l,b)$ model results resemble the most those for the shortest scale length $L=1.35$~kpc, for both the mode and median. This is consistent with the fact that  values of $L$ shorter than $1.35$~kpc are the most common in the BJ18 model (see their Fig.~1).

For the \emph{mode} $r_{\gal}$, the best performance is consistently obtained with the $L=2$~kpc estimator, which is effectively unbiased up to $\sim14.5$~kpc and showing the smallest bias ($\sim0.5$~kpc) at $15.5$~kpc. From $r_{\gal}>13.$~kpc, the short scale estimators $L=1.35$~kpc and $L$(BJ18) show a larger bias than for $L=2$~kpc ($\sim0.3$~kpc), half the corresponding bias expected for the median ($\sim0.6$~kpc). Therefore, for the $0.5$~kpc bins used in our analysis, the results for these two short scale length priors should not differ too much from those of the $L=2$~kpc estimator when using the mode.

The threshold radius of $\sim12-13$~kpc that marks the change in behaviour for the bias depends upon the selected tracer, as this choice will set a particular dependence of the parallax errors with distance, via the intrinsic magnitude, the colour and the distance modulus. Compared to the RC stars used in the tests presented here, we expect the threshold radius to be slightly larger for OB stars, around $16$~kpc, since these stars are intrinsically brighter on average, based on similar tests conducted in \citet{Abedietal14} using the reciprocal of the parallax as an estimator. There will also be an additional contribution to the $r_\gal$ bias due to contaminant stars, which will have a different error distribution. However, this contribution is small, $<10\%$ of the stars in either of our samples (see discussion in \ref{app:app_HR}) and largely due to dwarf stars. They affect mostly the radial bins at $r_\gal<10$~kpc and contribute at most $\sim0.15$~kpc, in addition to the bias shown in Fig.~\ref{f:RtruDist_L_medmod_DR2}, at any given radius.

As discussed in \citet{Luri2018} the best estimator depends upon the particular choice of the sample and its specific parallax error distribution. Our results show that there is no single estimator, of those considered here, that simultaneously outperforms all the others at all distance ranges and with all statistics. Overall, we find that for GaiaDR2 errors and $f_{obs}<0.5$, the mode as the estimator with a prior scale length $L=2$~kpc shows the best performance. 

\subsection{The use of \Gaia mock catalogues}
\label{sec:mocks}
In this work we make use of \Gaia mock catalogues in order to test the capabilities of the methods described above and the possible effects of \Gaia selection function and astrometric errors in the characterisation of the Galactic warp. Details about the warp models, the coordinate transformations and the generation of mock catalogues are given in Appendices~\ref{app:models}, \ref{app:veltrans} and \ref{s:test_scenarios}, respectively.

We generate three \Gaia mock catalogues, one to test the null hypothesis, i.e. a catalogue with no warp, the other two with imposed warp models, namely the Sine Lopsided and the S Lopsided model. We first generate a set of disc Red Clump initial conditions in a flat disc as in \citet{Abedietal14,RomeroGomez2015}. This set of particles will form our null hypothesis. We then generate a test particle simulation by integrating the initial conditions using the same integration strategy as in \citet{Abedietal14} using the Sine Lopsided, a warp model that allows the test particle simulation to reach statistical equilibrium. On the contrary, we cannot ensure statistical equilibrium for the S Lopsided models, i.e. we cannot guarantee that positions and velocities of a test particle simulation using the S Lopsided model will inherit those established by the model. We thus show the results of the random realization of particles, i.e. we apply the velocity transformation described in Appendix~\ref{app:veltrans} to the initial flat conditions. The free parameters defining the warp disc are taken from \citet{Amores17} for a mean age population of $2.5$~Gyr. 

Finally, we generate a GaiaDR2 mock catalogue up to magnitude G = 20 with the extinction model from \citet{Drimmel2003} and using the prescription of the astrometric, photometric and spectroscopic formal errors for Data Release 2\footnote{A fortran code to generate the Gaia errors is provided in \texttt{https://github.com/mromerog}}. We simulate errors for a mission time of $22$ months and, for the bright stars (G<13), we include a multiplicative factor of $3.6$ to the error in parallax to match the distribution of uncertainties as a function of the G magnitude observed in GaiaDR2 data. The final mock catalogues have more than 2 million Red Clump star particles.

\section{Selection of the two working samples in GaiaDR2}
\label{sec:sample_selection}

We are interested in selecting from the GaiaDR2 catalogue two samples characterised by being intrinsically bright, in order to reach the outermost parts of the disc, and with different age ranges, in order to assess whether the structural and kinematic properties of the warp depend on the age of the population. We select in the HR diagram of a young population formed by mainly upper main sequence stars or OB-type stars (hereafter, OB sample) and of an older population composed of all the stars in the red giant branch (hereafter, RGB sample). From GaiaDR2\footnote{We use the GaiaDR2 catalogue within the Big Data infrastructure known as the Gaia Data Analytics Framework (GDAF) cluster in the Universitat de Barcelona \citep{Tapiador2017}, firstly used for a scientific purpose in \citet{Mor2018}. It runs on Apache Spark (https://spark.apache.org/), which is an engine coming from business science suited to deal with large surveys.} we select stars up to magnitude G=20, with an available parallax measurement and with $f_{obs}<0.5$ (as discussed in the previous section), i.e. an absolute value of the relative error in parallax less than 50$\%$ (i.e., we keep stars with negative parallaxes). This first cut reduces the GaiaDR2 sample to $383,510,799$ sources. 

For these stars, we compute the distance using a Bayesian estimator with an exponentially decreasing space density prior with scale-length $L=2$~kpc, based on our analysis of the distance bias for different estimators (see Sect.~\ref{sec:distance_estimators}). Once we have derived the distances, and in order to reduce the computational cost of the process, we make a second cut which consists in removing cool main sequence stars from the sample. To do that, we select stars with $M_G^{\prime}$ < 1.95*($\bp$-$\rp$)+2., following the extinction line, where $M_G^{\prime}$ is the absolute G magnitude of the star uncorrected for extinction; $M_G^{\prime}$ is given by $M_G^{\prime}=G-5\log10(d)+5$; and ($\bp$-$\rp$) is the observed colour. More details are given in Appendix~\ref{app:app_HR}. This second cut reduces the sample to $86,814,618$ sources. Since we want to characterise the Galactic warp, we only keep stars with galactocentric spherical distance $r_\gal>7$~kpc and galactocentric cylindrical distance $R_\gal<16$~kpc\footnote{The LonKin methods works in galactocentric cylindrical bins, while the PCM works in galactocentric spherical bins.}. This second cut reduces the sample to $45,349,864$ stars. For these stars, we compute the V-band absorption using the Drimmel extinction model \citep{Drimmel2003} with re-scaling factors. Using $A_V$ and ($\bp$-$\rp$), we compute $A_G$ and the reddening E($\bp$-$\rp$) according to a polynomial fit (J.M. Carrasco, private communication). We refer to this sample as the "All-Stars" sample. Finally, the two tracers are selected as in \citet{Katz2018}: 
\begin{enumerate}
\item Young population, OB sample:
\begin{equation}
\begin{aligned}
M_G &< 2. \\ 
(\bp - \rp)_0&<0.
\label{eq:selectionOB}
\end{aligned}    
\end{equation}

\item RGB sample:
\begin{equation}
\begin{aligned}
M_G &< 3.9 \\ 
(\bp-\rp)_0&>0.95
\label{eq:selectionRGB}    
\end{aligned}    
\end{equation}

\end{enumerate}
where $M_G=M_G^{\prime}-A_G$ and ($\bp$-$\rp$)$_0$=($\bp$-$\rp$)-E($\bp$-$\rp$). The OB sample consists of $1,860,651$ stars and the RGB sample has $18,008,025$ stars. This RGB sample is a mixture of stars with ages in the range $3-7$~Gyr, while the OB sample typically is formed by young stars with age $\le 1$~Gyr \citep[e.g.][]{Robin12} . Using the Gaia Universe Model Snapshot \citep{Robin12} with GaiaDR2 simulated errors, we have  estimated the contamination in the OB and RGB samples to be $9\%$ and $8\%$, respectively (see Appendix~\ref{app:app_HR}).

Figure~\ref{fig:hist} shows the resulting distributions in G magnitude and heliocentric distance, $d$, for both working samples. We emphasise here that both samples reach up to magnitude $20$ and they extend well beyond the solar neighbourhood, having $1.1\times 10^6$ OB type stars and $13.6\times 10^6$ RGB stars up to $d=5$~kpc.

\begin{figure}
\begin{center}
 \includegraphics[width=\columnwidth]{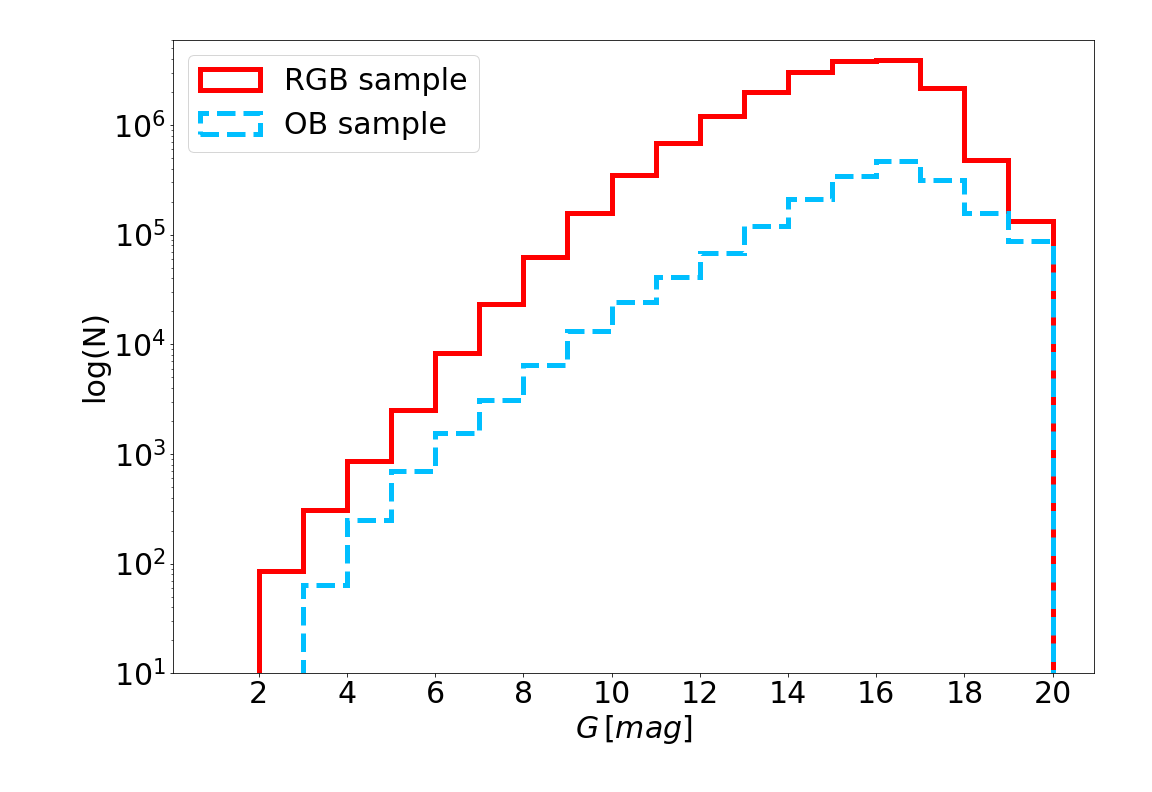}
 \includegraphics[width=\columnwidth]{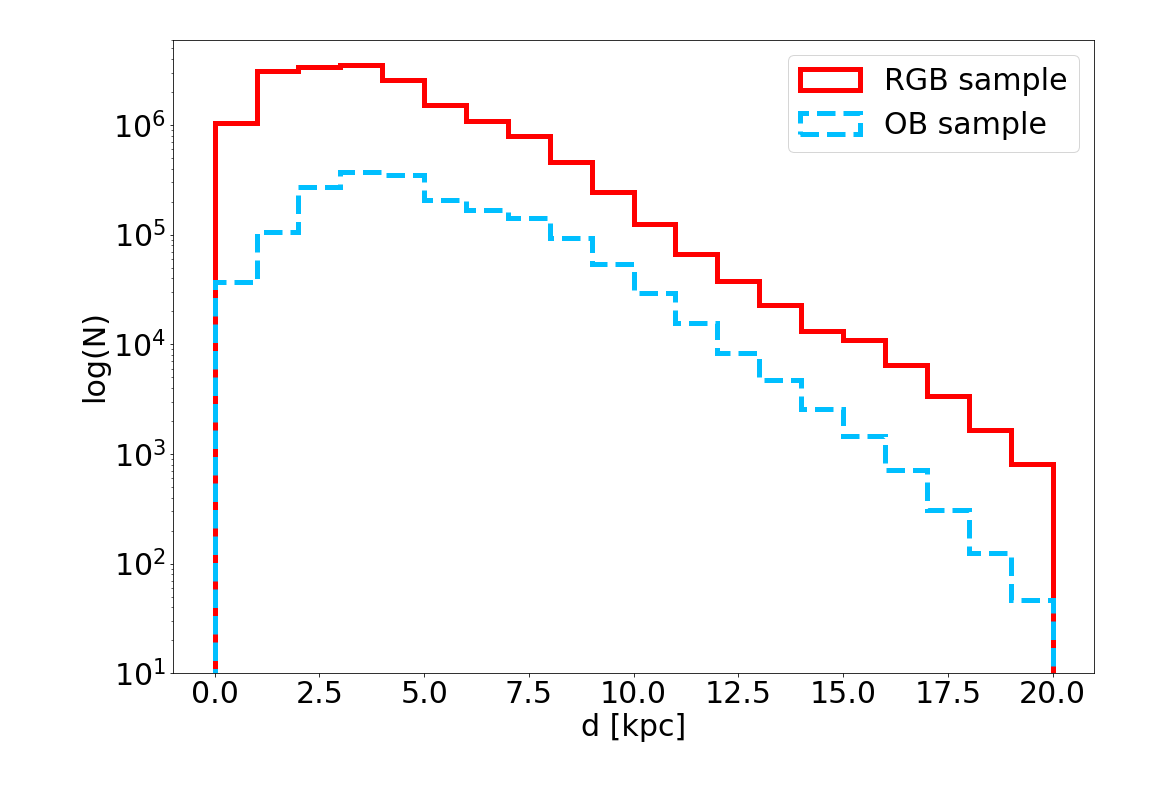}
 \caption{Distributions in magnitude G (top panel) and heliocentric distance (bottom panel) of the two working samples.}
\label{fig:hist}
\end{center}
\end{figure}

\section{The spatial distribution of the OB and RGB samples}
\label{sec:spatial}

In this section, we analyse the spatial distribution of both working samples in order to detect the signature of the warp. We apply a coordinate transformation to the galactocentric cartesian frame and we study the structural characteristics of both samples. Throughout this work, we use a right-handed galactocentric reference frame. Its X-axis points along the Sun-Galactic centre line, away from the Sun. The Y-axis is orthogonal to the X-axis and on the Galactic plane in the direction of Galactic rotation. The Z-axis points towards the North Galactic pole. We also use cylindrical coordinates $(R,\theta, Z)$ with $R$ the Galatocentric distance and the azimuthal angle $\theta$ measured on the Galactic plane starting from the positive X-axis and towards the positive Y-axis. To perform these coordinate transformations from the GaiaDR2 data, we adopt a distance of the Sun to the Galactic centre of $R_{\odot}=8.34$~kpc \citep{Reid2014} and the height of the Sun with respect to the Galactic plane of $Z_{\odot}=0.027$~kpc \citep{Chen2001}. We also adopt the circular velocity at the solar radius from \citet{Reid2014} of $V_c(R_{\odot})=240$~\kms \footnote{We use these values for consistency and continuation of the \citet{Katz2018} work. We have checked the more recent values for the Solar radius and circular velocity at the solar radius from \citet{Grav2018} do not change at all the results of the paper.}. We adopt the peculiar velocity of the Sun with respect to the local standard of rest of $(U_{\odot},V_{\odot},W_{\odot})=(11.1\substack{+0.69 \\ -0.75},12.24\substack{+0.47 \\ -0.47},7.25\substack{+0.37 \\ -0.36})$~\kms \citep{Schoenrich2010}.

\subsection{Peculiar features in the stellar density distribution}
\label{sec:spatial_maps}
\begin{figure}
\begin{center}
 \includegraphics[width=\columnwidth]{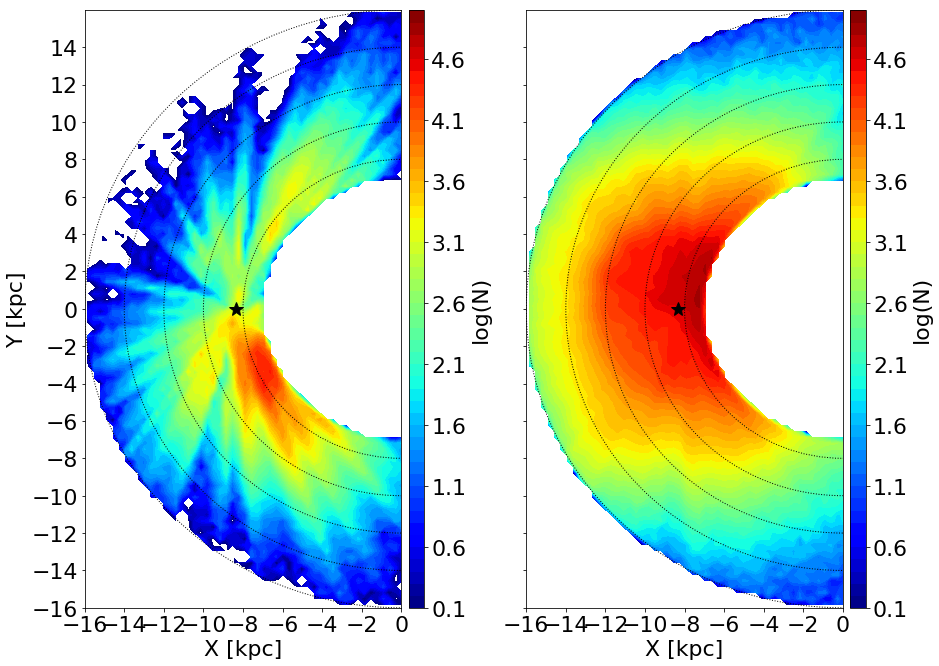}
 \caption{Two dimensional stellar density map of the two working samples, namely OB sample (left column) and RGB sample (right column). The size of the bins are $0.32$~kpc $\times\, 0.64$~kpc for the OB sample and $0.16$~kpc $\times\, 0.32$~kpc for the RGB sample. We mark the Sun position with a star and the black dotted lines show circles at different galactocentric radii, R=$8$, $10$, $12$, $14$ and $16$ kpc. The colour bar is in log-scale and common for both panels. The Galaxy rotates clockwise.}
\label{fig:projdens}
\end{center}
\end{figure}

In Fig.~\ref{fig:projdens} we plot the stellar density map of the two samples in the (X,Y) cartesian galactocentric projections. In general, both populations show the radial features characteristic of the high extinction lines (under-densities), which point to the presence of dust clouds in the direction of the line-of-sight, and of the error in distance of a given object (shown as elongated over-densities, also known as the finger-of-god effect). The effect of dust clouds is visible in both populations (OB and RGB samples), being the most prominent the ones we find in the directions: $75\degr<l<85\degr$, $130\degr<l<150\degr$, $180\degr<l<190\degr$ and $250\degr<l<270\degr$, well in agreement with the Drimmel extinction map \citep{Drimmel2003}.

Focusing on the stellar density map of the OB sample (see left panel of Fig.~\ref{fig:projdens}), the over-densities we observe clearly trace some known star forming regions. Zooming in at heliocentric distances closer than one kpc, we could see the Orion, Vela and the Cygnus star forming regions, well described in \citet{Zari2018}. At heliocentric distances between $1<d<3$~kpc, we see the Cassiopeia region ($120\degr<l<130\degr$), the Outer Spur ($230\degr<l<250\degr$) and the Carina star forming region ($280\degr<l<300\degr$). In the latter, it is where we have a larger concentration of OB stars. This over-density is well traced by the early type stars, including HII regions \citep[e.g.][]{MolinaLera2016} or the classical Cepheids \citep{Skowron2018}, and it is often associated to the Carina-Sagittarius spiral arm of gas and dust \citep[e.g.][]{Martos2004}. We observe that this Carina region over-density contributes to the galactocentric rings $8<R<10~$kpc in the study of the Galactic warp.

\begin{figure}
\begin{center}
 \includegraphics[width=\columnwidth]{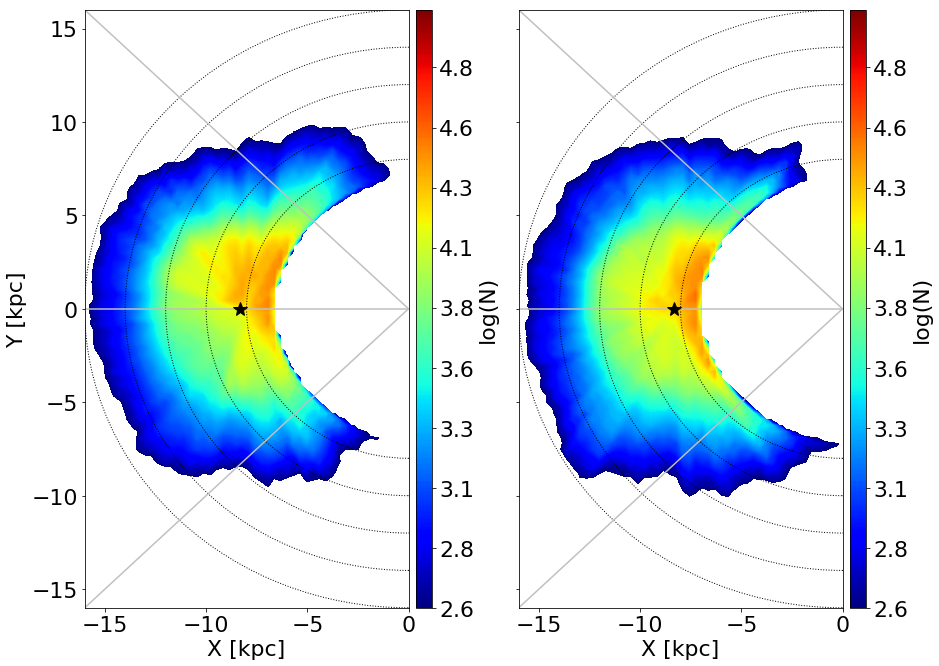}
 \caption{Two dimensional projections of the RGB sample. Left panel: Z>0. Right panels Z<0. The size of the bins are $0.16$~kpc $\times\, 0.32$~kpc for the RGB sample. The colour mapping scales logarithmically with number counts $N$. We mark the Sun's position with a star and the gray solid lines show the galactocentric azimuths: $135\degr$, $180\degr$, and $225\degr$. The black dotted lines show circles at different galactocentric radii, R=$8$, $10$, $12$, $14$ and $16$ kpc. }
\label{fig:projZupZdown}
\end{center}
\end{figure}

In the right panel of Fig.~\ref{fig:projdens}, we plot the stellar density map corresponding to the RGB sample. Although it apparently looks more homogeneous than the map for the OB stars, it presents a clear asymmetry in stellar density with respect to the centre--anticentre line. We show in detail this feature in Fig.~\ref{fig:projZupZdown}, where we separate the RGB stars that are located above (left panel) and below (right panel) the Galactic plane. Although some of the under-densities visible in both panels have the characteristic radial trend of high extinction lines, this significant over-density region at galactocentric azimuths $135\degr<\theta<180\degr$ is clearly visible in both hemispheres, and not expected in an axisymmetric Galactic potential. This over-density will require further detailed analysis in the future. 
Combining the information in Figs.~\ref{fig:projdens} and \ref{fig:projZupZdown}, we conclude that the stellar density maps of the young OB and old RGB samples clearly show opposite over-densities with respect to the centre--anticentre line at galactocentric radial bins between $8<R<12~$kpc.

\subsection{Spatial signature of the Galactic warp}
\label{sec:spatial_warp}
\begin{figure}
\begin{center}
 \includegraphics[width=\columnwidth]{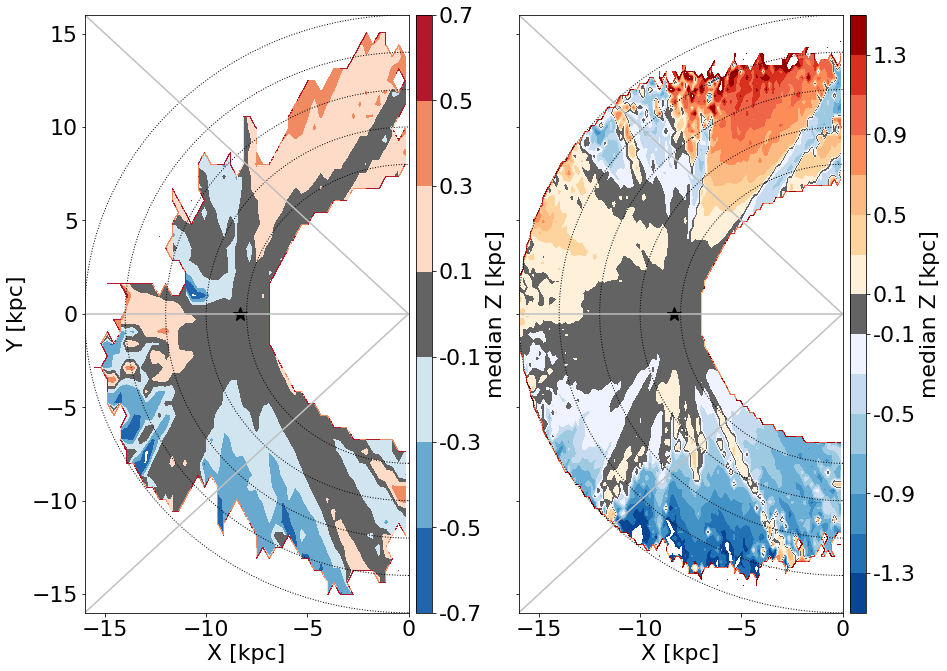}
 \caption{Two dimensional projections of the median Z for the two working samples, namely OB sample (left column) and RGB sample (right column). The size of the bins are $0.32$~kpc $\times\, 0.64$~kpc for the OB sample and $0.16$~kpc $\times\, 0.32$~kpc for the RGB sample. We only plot the bins with at least $15$ sources. We mark the Sun position with a star, the gray solid lines show the galactocentric azimuths: $135\degr$, $180\degr$, and $225\degr$, and the black dotted lines show circles at different galactocentric radii, R=$8$, $10$, $12$, $14$ and $16$ kpc. The color scale is different in each sample.}
\label{fig:medZ}
\end{center}
\end{figure}

\begin{figure}
\begin{center}
 \includegraphics[width=\columnwidth]{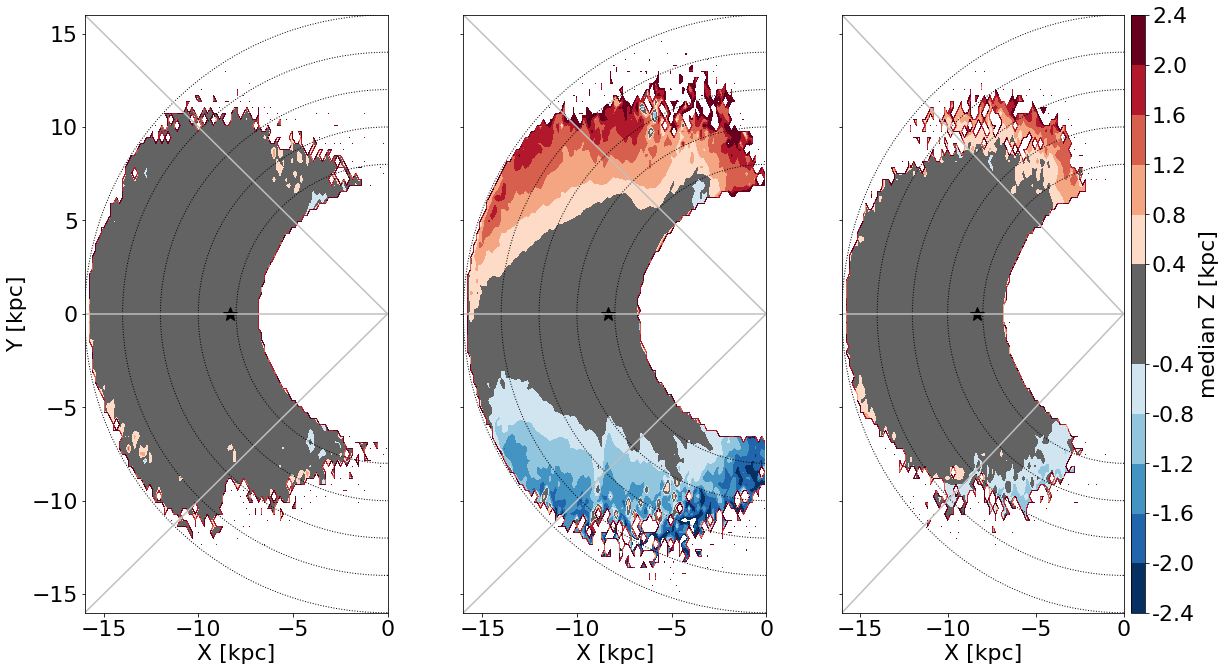}
 \caption{Two dimensional projections of the median Z for the three mock catalogues, namely the null hypothesis (flat disc) (left), the Sine Lopsided warp disc (middle) and the S Lopsided warp disc (right). The size of the bins are $0.16$~kpc $\times\, 0.32$~kpc. We mark the Sun position with a star, the gray solid lines show the galactocentric azimuths: $135\degr$, $180\degr$, and $225\degr$, and the black dotted lines show circles at different galactocentric radii, R=$8$, $10$, $12$, $14$ and $16$ kpc. }
\label{fig:medZmock}
\end{center}
\end{figure}

We now focus on the stellar signature of the warp detected at larger heliocentric distances by plotting the median Z for each of the samples (see Fig.~\ref{fig:medZ}). Both samples show a clear positive median in Z at galactic azimuths $90\degr<\theta<135\degr$ and a negative median at $225\degr<\theta<270\degr$, well in agreement with the expected behaviour of the Galactic warp \citep[e.g.][]{Drimmel2000,Levine2006}. For intermediate azimuths, $135\degr<\theta<225\degr$, where the signature of the line-of-nodes lies, the data shows a high degree of complexity, with significant differences between the two populations. By comparing with the warped mock catalogues with a Sine Lopsided warp model and a straight line-of-nodes in the Galactic centre -- anti-centre direction (see middle panel of Fig.~\ref{fig:medZmock}), we could associate a stripe with null median Z towards the anti-centre direction as the possible location of the line-of-nodes of the Galactic warp. We see that the extinction can modify the shape of the line-of-nodes so any solid conclusion about the position of the line-of-nodes requires a detailed correction for extinction.
The RGB sample with more than 18 million stars, shows a null median Z stripe inclined with respect to the Sun--anticentre direction towards $\theta\sim 180-200\degr$. The fact that the stripe is curved and with galactic azimuth growing with galactocentric radius could induce to think that the line-of-nodes is twisted, though as mentioned above, the position could also be affected by the extinction. We emphasise that these results do not agree with previous works. As an example, \citet{Momany2006} classically placed the line-of-nodes at $\theta<180\degr$. Regarding the young population, our OB sample presents a clumpy distribution of median Z, without a clear trace of the line-of-nodes. Some recent studies using classical Cepheids, place the line-of-nodes of the warp also at $\theta<180\degr$ \citep{Skowron2018,Chen2019}. The Cepheids population is very young, so we would expect the OB sample to be comparable to the \citet{Skowron2018} and \citet{Chen2019} results. However, as mentioned above, the clumpy structure of the median Z we find for this young population, does not allow us to support or rule out their conclusion.

For comparison, in Fig.~\ref{fig:medZmock} we show the two dimensional projection of the median Z for the three mock catalogues, namely with a flat disc -- null hypothesis, with a Sine Lopsided warp model and the S Lopsided warp model, from left to right, respectively. In the left panel, the disc imposed is flat and therefore, the median Z is around zero, as seen in the large extent of the disc. There are small regions in which the median deviates slightly from zero because of the effect of the extinction and large astrometric errors. In the case of a mock catalogue including the effect of the Galactic warp (middle and right panels of Fig.~\ref{fig:medZmock}), we see the expected median of Z for such warped model, including some line-of-sight features, which correspond to lines with high astrometric errors, namely $l\sim 50\degr$ and $l\sim 300-310\degr$. These lines are also seen in GaiaDR2 data (see Fig.~\ref{fig:medZ}), but the effect of the extinction or GaiaDR2 errors do not mask the spatial signature of the Galactic warp. In the middle panel, for the SineLop model, we note the gray stripe that should correspond to the line-of-nodes bends in the anti-centre direction towards galactic azimuths $\theta>180$\degr, as we also see in the GaiaDR2 data and mentioned above. On the other hand, in the SLop model (right panel), the gray stripe is broader as expected by construction of the SLop model (see Fig.~\ref{fig:models} in appendix~\ref{app:models}), thus it is not limited to a small region along the line-of-nodes.

\begin{figure}
\begin{center}
 \includegraphics[width=\columnwidth]{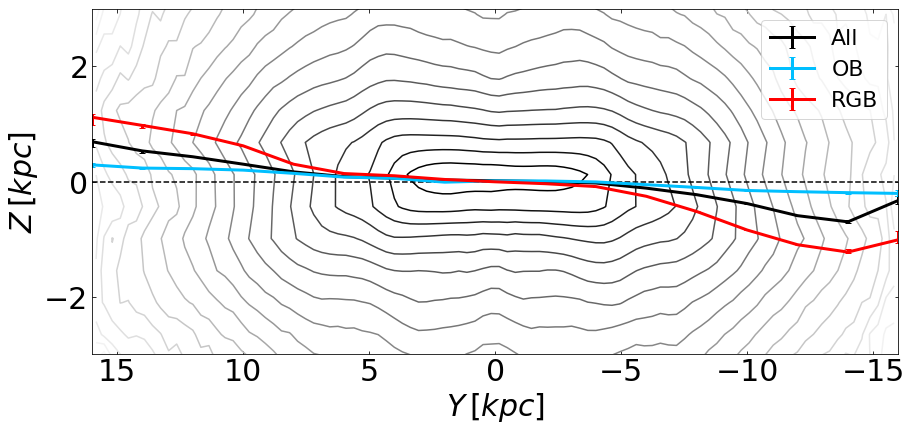}
 \caption{Contour density plots in the (Y,Z) projection of the All-Stars sample in gray scale. Solid lines show the median for the three samples: in black, blue and red, the All-Stars, OB and RGB samples, respectively. Note that the Y-axis is inverted. The line shows the median of the particles in bins of $2$~kpc and the error bars show the lower and upper $1\sigma$ uncertainty in the position of the median.}
\label{fig:projZ}
\end{center}
\end{figure}

Finally, in Fig.~\ref{fig:projZ} we plot the density contours in the (Y,Z) projection. In this projection, the X-axis is perpendicular to the plot and we expect to see the up/down side of the warp towards the Y-positive/Y-negative axis, respectively. We show the median of the three samples, namely All-Stars (black), OB (blue) and RGB (red) samples. As expected, the median of the density in the (Y,Z) projection for the three samples is not flat, but it describes a warped profile. More important, the GaiaDR2 data definitively confirms that the amplitude of the warp is increasing with the age of the population. 

The altitude of the warp at $R=14$~kpc for the OB and RGB samples is given in the top rows of Table~\ref{tab:Zlit}, where we also provide values given in other works found in the literature, for comparison. Our results show that the altitude of the Galactic warp at $R=14$~kpc is of the order of $0.2$~kpc and $1.0$~kpc for the OB and RGB samples, respectively. These values are in agreement with those obtained by \citet{Drimmel2001}, \citet{LopezCorredoira2002} using a Red Clump population, also by \citet{Reyle2009} using 2MASS star counts also dominated by the red giant population and, finally, with those of \citet{Marshall06} traced by dust. We also include the results derived using pulsars \citep{Yusifov2004} as it is generally accepted that progenitors of pulsars are young OB stars. The estimated altitude of the warp using the OB sample is smaller than that determined by the pulsars. Even though our results can be biased by the effects of the extinction, our results suggest that \citet{Amores17} values for all age populations are over-estimated. Finally, we also want to mention that the RGB sample shows an inflection point at around $R\sim 13-14$~kpc at the down side, similar to the one reported by \citet{Levine2006} when tracing the warp in the gas. Further work is required to analyse why such different tracers present a similar feature in this region.

Our RGB sample shows only a small degree of asymmetry in the sense that $|Z(up)|\lesssim |Z(down)|$, in agreement with \citet{Marshall06} and opposite to the values reported by \citet{Amores17}, who, in the age range $2-7$~Gyr, always found that $|Z(up)|>|Z(down)|$.

To conclude, the stellar warp is highly dependent on the age of the tracer population. Whereas no solid conclusion can be established from the OB sample, the huge amount of data of the GaiaDR2 RGB sample suggests a slightly lopsided in the sense $|Z(up)|\lesssim |Z(down)|$, with a line-of-nodes of the Galactic warp twisted towards galactocentric azimuths $\theta\sim 180-200\degr$ at $R>12$~kpc.

\begin{table}[]
    \centering
    \begin{tabular}{|c|c|c|c|}
    \hline
    Source                       & $Z_{up}$ & $Z_{down}$ & Wavelength\\
    \hline
    \hline
    OB (this work)        & 0.23 & -0.19  & Gaia optical\\
    \hline
    RGB (this work)       & 0.97 &  -1.22 & Gaia optical\\
    \hline
    D01          & 1.34 & $\sim$ & COBE/DIRBE NIR \& FIR\\
    \hline
    LC02  & 1.23 & $\sim$ & 2MASS NIR\\
    \hline
    Y04  & 0.62   &  -0.58  &  ATNF pulsars  \\
    \hline
    M06 (dust)    & 0.74 & -0.68 & 2MASS NIR \\
    \hline
    R09            & 0.50 & $\sim$ & 2MASS NIR\\
    \hline
    A17 (0.5Gyr)    & 3.3  & -0.4  & 2MASS NIR \\
    \hline
    A17 (2.5Gyr)    & 2.1  & -1.2  & 2MASS NIR  \\
    \hline
    A17 (4.0Gyr)    & 2.0  & -1.7 & 2MASS NIR  \\
    \hline
    A17 (6.0Gyr)    & 3.9  & -2.7 & 2MASS NIR  \\
    \hline
    \end{tabular}
    \caption{Values of the warp up ($Z_{up}$) and down ($Z_{down}$) altitude with respect to the Galactic plane at galactocentric radius $R=14$~kpc expressed in kpc. If the authors report an average value or find no difference between the up and down side, we specify it with the symbol $\sim$ in the $Z_{down}$ column. We refer to \citet{Drimmel2001} as D01, \citet{LopezCorredoira2002} as LC02, \citet{Yusifov2004} as Y04 using ATNF Pulsar catalogue \citep{Manchester2005},  \citet{Marshall06} as M06,  \citet{Reyle2009} as M09,  \citet{Amores17} as A17.}  For the \citet{Amores17} we indicate in parenthesis the mean age of the population.
    \label{tab:Zlit}
\end{table}

\section{The kinematic signature of the Galactic warp}
\label{sec:kine}
In this section, we analyse the kinematic distribution of both working samples using three strategies. We start by projecting the median proper motion distribution into the Galactic plane (Sect.~\ref{sec:kine_maps}) and then by applying the two methods described in Sect.~\ref{sec:methods}, namely the LonKin (Sect.~\ref{sec:lonkin_data}) and the \ngc~PCM methods (Sect.~\ref{sec:ngc_data}), which extract and quantify the overall signal of the warp using both position and kinematic information. We show simplified theoretical expectations for three different warp models\footnote{The warp models used are described in detail in appendix~\ref{app:models}.} as a guide on how to interpret these results.

\subsection{The kinematics in the (X,Y) plane}
\label{sec:kine_maps}

\begin{figure}
\begin{center}
 \includegraphics[width=\columnwidth]{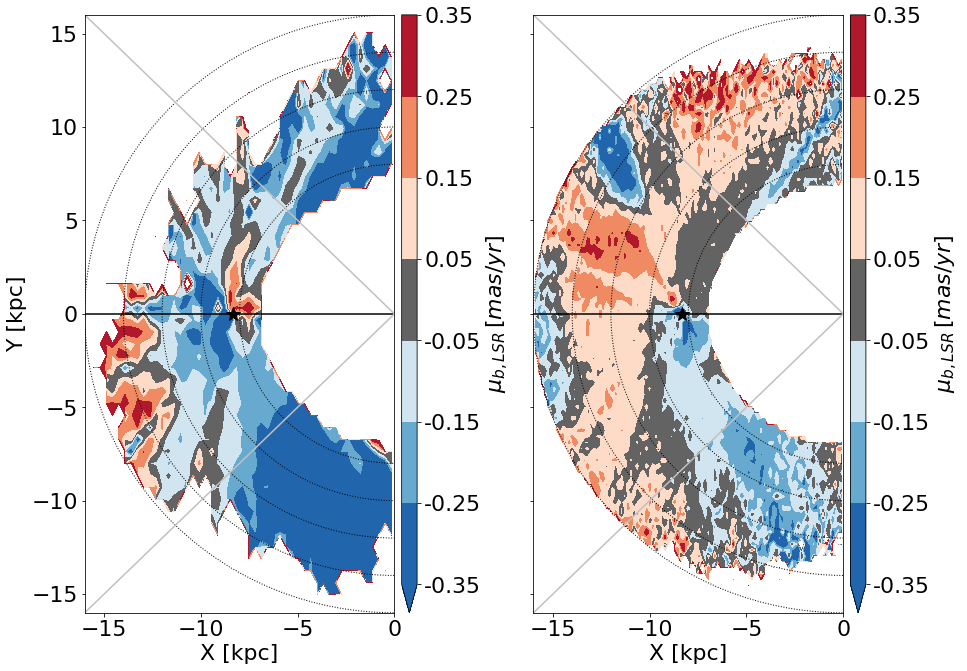}
\caption{Two dimensional maps of the median proper motion in the latitude direction corrected for the Sun's peculiar motion, $\mu_{b,LSR}$ of the two working samples, namely OB sample (left) and RGB sample (right). The black dotted lines show circles at different galactocentric radii, R=$8$, $10$, $12$, $14$ and $16$ kpc. The size of the bins are $0.32$~kpc $\times\, 0.64$~kpc for the OB sample and $0.16$~kpc $\times\, 0.32$~kpc for the RGB sample. We only plot the bins with at least $15$ sources. The grey solid lines show azimuths $135\degr$ and $225\degr$, while the black horizontal line show the centre--anti-centre direction. The position of the Sun is shown with a black small star.}
\label{fig:2dmovprop}
\end{center}
\end{figure}

In Fig.~\ref{fig:2dmovprop}, we plot the median of the proper motion in latitude corrected for the Sun's peculiar motion projected into the (X,Y) plane and we analyse any substructures seen in this projection.

\begin{figure}
\begin{center}
 \includegraphics[width=\columnwidth]{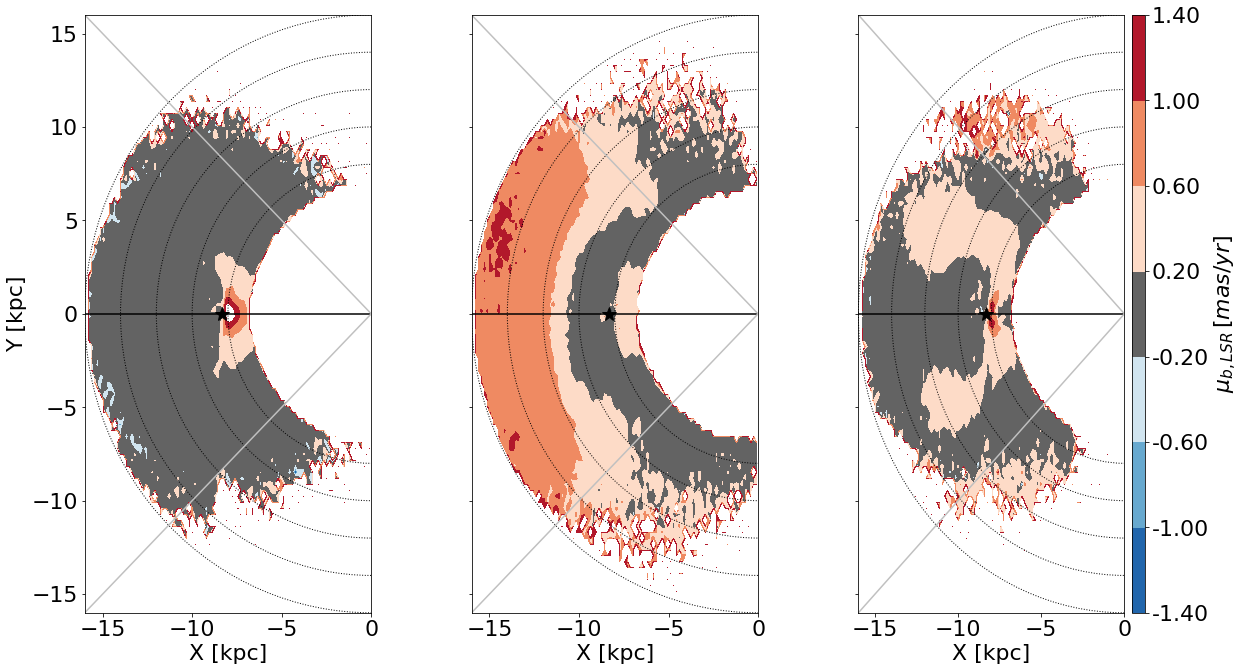}
 \caption{Two dimensional projections of the median $\mu_{b,LSR}$ for the three mock catalogues, namely the null hypothesis (flat disc) (left column) and the Sine Lopsided warp disc (middle column) and the S Lopsided warp disc (right column). The size of the bins are $0.16$~kpc $\times\, 0.32$~kpc. We mark the Sun position with a star, the gray solid lines show the galactocentric azimuths: $135\degr$, $180\degr$, and $225\degr$, and the black dotted lines show circles at different galactocentric radii, R=$8$, $10$, $12$, $14$ and $16$ kpc. }
\label{fig:medmubLSRmock}
\end{center}
\end{figure}

The median $\mu_{b,LSR}$ distribution shows some radial features in the kinematic map, such as those at l $\sim 55\degr,\,135\degr,\,200\degr,\,220\degr$. Such features are also present in \citet{Poggio2018}, with similar but less extended samples. By using the mock catalogues, we study whether we can attribute these radial features to effects of the extinction. In Fig.~\ref{fig:medmubLSRmock} we show the median $\mu_{b,LSR}$ distribution for the mock catalogues with flat disc (null hypothesis, left panel) and with a Sine Lopsided and S Lopsided warped disc (middle and right panels). As expected, the proper motion in latitude has a median around zero in the case of a flat disc. According to a simple symmetric warp model (see appendix~\ref{s:ss_model}), as stars are moving from the down-side at $\theta\sim270\degr$ to the up-side at $\theta\sim90\degr$, the expected distribution of the proper motion in latitude $\mu_{b,LSR}$ has maximum and positive values towards the anti-centre decreasing from there towards both sides \citep{Abedi2015}. In the case of a Sine Lopsided disc, the maximum of the proper motion is shifted towards the side of larger amplitude (the up side in the mock catalogues), while in the case of a S Lopsided disc the proper motion in latitude has a modulation, being almost zero along the line-of-nodes and maximum towards the up and down lines-of-sights. We can conclude that the effect of extinction or GaiaDR2 errors do not affect the predicted pattern. 

The pattern we observe in the two dimensional projections in the OB and RGB samples (see Fig.~\ref{fig:2dmovprop}) shows that, towards the anti-centre, the median $\mu_{b,LSR}$ has not a uniform positive trend, but shows modulated changes. Whereas the median $\mu_{b,LSR}$ of the OB sample have a transition from negative to positive values, the reverse is observed in the RGB sample. This is a clear first signature of the age dependency of the vertical motion. Its possible relation to a bending mode \citep{Katz2018} is discussed in Sect.~\ref{sec:discussion}.

These maps also allow us to detect any possible asymmetry with respect to the Sun--anti-centre line, which might be related to the lopsidedness of the Galactic warp. For the RGB sample (right panel of Fig.~\ref{fig:2dmovprop}) we detect a significant stripe with maximum positive $\mu_{b,LSR}$ towards galactic azimuths $\theta\sim 160-170\degr$. Even more, this shift of about $20-30\degr$ from the Sun--anti-centre line is also visible all through the map, making the null median values to be asymmetric (further discussed in Sect.~\ref{sec:lonkin_data}). For the OB sample (left panel of Fig.~\ref{fig:2dmovprop}), the kinematic structure is more clumpy, but again there is a clear shift with respect to the Sun--anti-centre line, with lower negative median $\mu_{b,LSR}$ towards the down side of the warp (l $\sim 270-310\degr$). 

Finally, we report the existence of a feature at l $\sim 100-120\degr$ at the outermost radii $R>12$ kpc in the RGB sample. We refer to it as the "blob". The lack of OB stars in this area does not allow us to know whether there exists a counterpart of this feature in the young population. We suggest that the "blob" is not related to the warp and we discuss it further in Sect.~\ref{sec:discussion}.

\subsection{The warp as seen by the LonKin method}
\label{sec:lonkin_data}

\begin{figure}
\begin{center}
\includegraphics[width=\columnwidth]{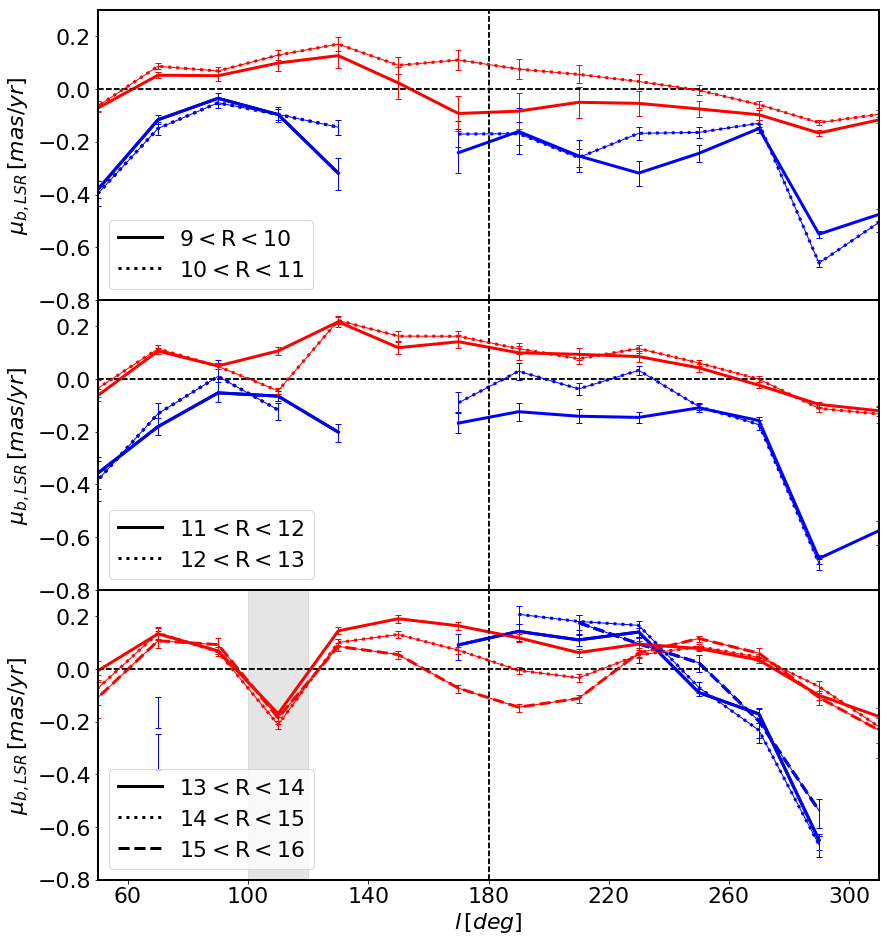}
\caption{LonKin method applied to the OB sample (blue lines) and RGB sample (red lines). We divide in three panels according to galactocentric rings, from the inner to the outer disc, from top to bottom. In each panel, different distance bins are indicated with solid, dotted and dashed lines consistently from nearest to farthest. The shaded vertical region in the bottom panel corresponds to the "blob" (see text for details). The line shows the median of the particles in bins of $20\degr$ and the error bars show the lower and upper $1\sigma$ uncertainty in the position of the median and they include in quadrature the uncertainty in the solar velocity.}
\label{fig:Lonkinsamples}
\end{center}
\end{figure}

Here we show the LonKin result for the OB and RGB, together with a toy theoretical prediction. Figure~\ref{fig:Lonkinsamples} shows the results for the OB sample (blue lines) and the RGB sample (red lines) in galactocentric radial bins of increasing distance from top to bottom panels. Only the longitude bins for which there are at least 300 stars are shown. Figure~\ref{fig:Lonkinsamples_theo} shows the model predictions for a random realisation of three warp models, described in detail in appendix~\ref{app:predictions}.

As already hinted in the previous section, GaiaDR2 data reveals clear differences between the two populations. As for the OB stars, the inner annuli show an almost flat trend with a median proper motion of about $-0.2$ \mas, while for the RGB, the median is slightly positive. In the middle annuli, the proper motion increases, as expected in a simple warp model, showing the OB stars a quite planar trend in the outer disc, from l=$90\degr$ to l=$270\degr$, and a sharp decrease towards the inner disc, whereas the RGB stars clearly show that the maximum in the proper motion is not achieved in the anti-centre direction, while it shows two local maxima at about l=$140\degr$ and l=$220\degr$,  the one in the second quadrant being slightly higher. As we move towards the outer annuli (bottom panel), the OB stars distribution becomes clearly asymmetric (see top left panel of Fig.~\ref{fig:projdens}), and the LonKin only provides information in the third and fourth quadrant, showing a maximum in the proper motion around l=$220\degr$. The proper motion for the RGB sample shows a clearly different trend from that of the OB stars. The trend that we could initially glimpse in the middle annuli, is clear here, with  the maxima moving slightly towards the inner disc. Also, the median proper motion in the anti-centre direction is almost zero for the radial bin $14<R<15$~kpc, and becomes negative in the outermost ring. We can also distinguish a significant local minimum at l$\sim 120\degr$, this corresponds to the "blob" already detected in the two dimensional maps shown above (see Fig.~\ref{fig:2dmovprop}). This minimum in proper motion in latitude could be related to a negative median Z at the same region (see right panel of Fig.~\ref{fig:medZ}) due to high extinction and the fact that the distribution of RGB stars is not homogeneous above and below the disc. It might also be caused by some type of substructure present in the Southern hemisphere.

\begin{figure}
\begin{center}
\includegraphics[width=\columnwidth]{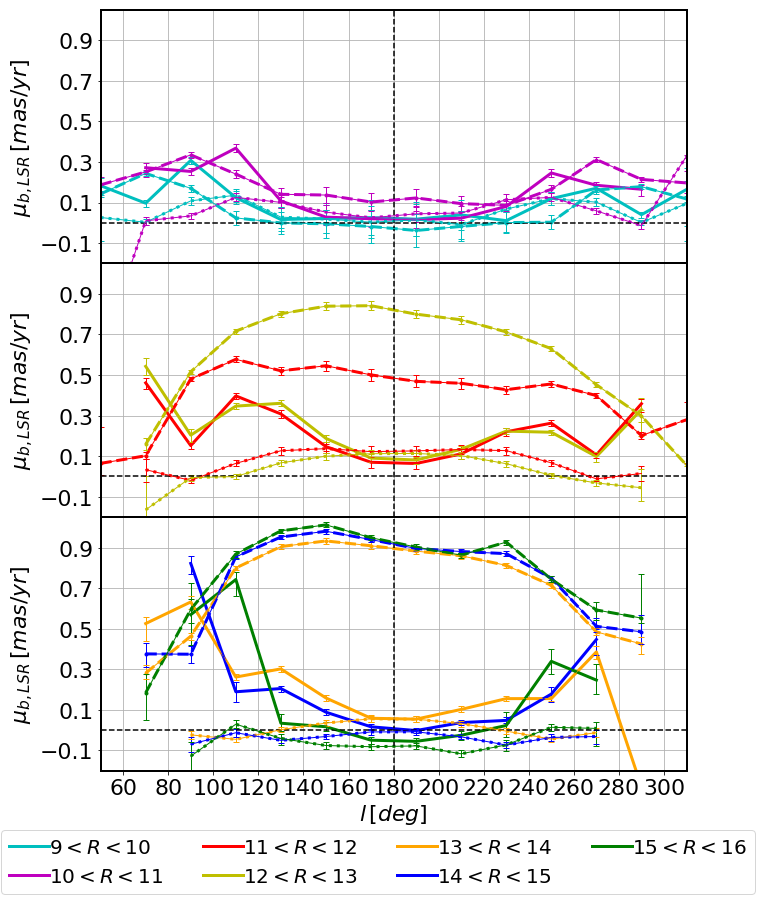}
\caption{LonKin method applied to the mock catalogues without warp (dotted lines), with the SineLop warp (dashed lines) and with the SLop warp (solid lines) models, including extinction, Gaia selection function and GaiaDR2 errors. The line shows the median of the particles in bins of $20\degr$ and the error bars show the lower and upper $1\sigma$ uncertainty in the position of the median and they include in quadrature the uncertainty in the solar velocity.}
\label{fig:Lonkinsamples_mock}
\end{center}
\end{figure}

For comparison, in Fig.~\ref{fig:Lonkinsamples_mock} we show the trend in proper motion  predicted by the three mock catalogues, namely the flat disc (dotted lines), the SineLop (dashed lines) and the SLop (solid lines) warped disc models. As expected, the proper motion in latitude for a flat disc is about null, while for a SineLop warped disc it increases towards the outer parts and it is not symmetric with respect the line-of nodes ($l=180$~\degr), and for the SLop warped disc it shows a modulation being around zero towards the line-of-nodes and it has two relative maxima towards the largest amplitudes. As mentioned in Sect.~\ref{sec:mocks}, the SineLop and SLop models are fixed using the free parameters given by \citet{Amores17} with $<age>=2.5$~Gyr., even though we have found these to be over-estimated (Sect.~\ref{sec:spatial}). The amplitude of the curves in Fig.~\ref{fig:Lonkinsamples} supports our findings from Sect.~\ref{sec:spatial} that the warp has a low amplitude. The fact that the proper motion for the RGB data has a double-peak tendency and that it becomes negative towards the anti-centre at large galactocentric distances ($14<R<15$~kpc) indicates that an S Lopsided model would be a right choice as a starting model (see further discussion in Sect.~\ref{sec:discussion}).

The LonKin method also allows us to estimate the initial radius of the warp. We find that the initial radius for the OB sample is $12<R<13$~kpc, while for the RGB sample is $10<R<11$~kpc. We again confirm an age dependence of the warp characteristics, as already shown in Sect.~\ref{sec:spatial}.

\begin{figure}
\begin{center}
 \includegraphics[width=0.9\columnwidth]{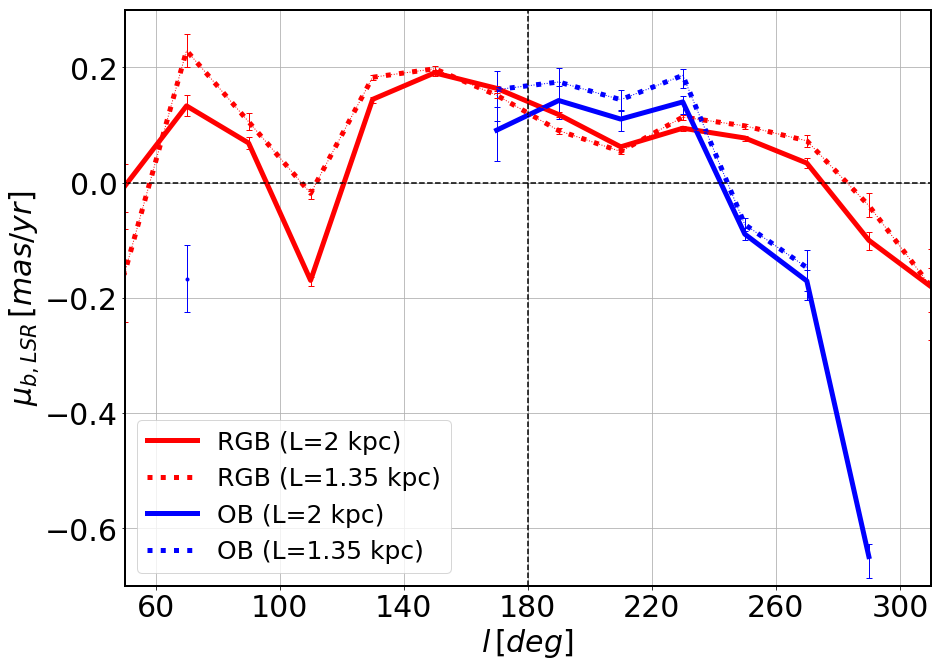}
\caption{Comparison of results for the LonKin methods for the radial bin $R=13-14$~kpc obtained for two different prior scale-lengths, L=$2$ kpc (solid lines) and L=$1.35$ kpc (dotted lines) in red for the RGB sample and in blue for the OB sample.}
\label{fig:OB_RGB_lonkin_robust_dist}
\end{center}
\end{figure}

In Fig.~\ref{fig:OB_RGB_lonkin_robust_dist} we check whether the results of applying the LonKin method shown in Fig.~\ref{fig:Lonkinsamples} depend on the choice of the distance estimator. We show the results of applying the LonKin method to the radial annuli $13<R<14$ with $L=2$~kpc (solid line) and with $L=1.35$~kpc (dotted line) for the OB (blue line) and the RGB (red lines) samples. Note that, qualitatively, both sets of curves are consistent within their respective error bars, which overlap in most longitude bins. Therefore, our results are robust against the choice of the distance estimator.

\subsection{\ngc~signature of the warp}
\label{sec:ngc_data}

\begin{figure*}
\begin{center}
 \includegraphics[width=1.0\columnwidth]{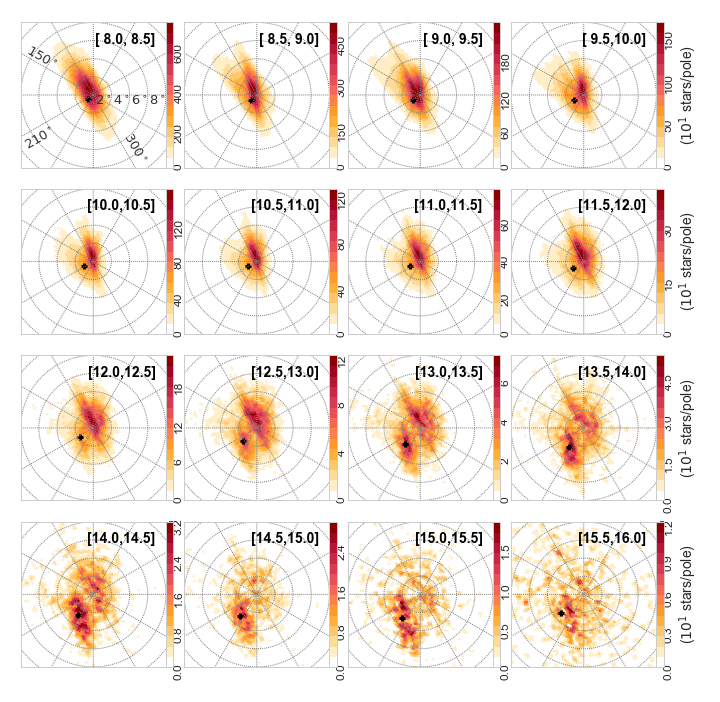} \includegraphics[width=1.0\columnwidth]{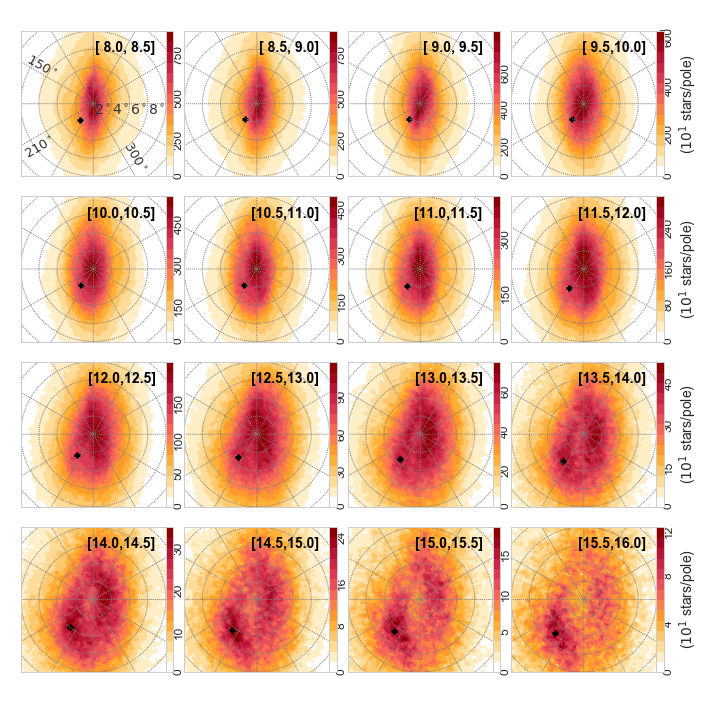}
\caption{\ngc~PCMs for the OB (left panels) and RGB samples (right panels). The corresponding $r_\gal$ range is indicated at the top right of each panel. The cross indicates the centroid (weighed by pole counts) of the main off-pole peak of the PCM. The maps were produced using a grid spacing and tolerance of $0\fdg25$. The colour scales linearly with star counts, indicating larger values with darker colours. In all panels, dotted circles and radial lines correspond respectively to parallels at co-latitudes increasing by $2\degr$ and meridians at azimuths every $30\degr$.}
\label{fig:PCM_data}
\end{center}
\end{figure*}

Here we show the \ngc~result for the OB and RGB, together with a toy theoretical prediction.
The \ngc~PCMs for the OB and RGB samples are shown in the left and right panels of Fig.~\ref{fig:PCM_data}, respectively, for $r_\gal$ bins of $0.5$~kpc width. Qualitatively the two sets of PCMs for both tracers have a similar behaviour: there is a peak at the north Galactic pole, i.e. the centre of the PCM, that remains fixed at all $r_\gal$ and another peak in the lower-left quadrant of the PCM, whose co-latitude increases as $r_\gal$ does. This latter peak is the clear signature of a warped disc whose amplitude increases with radius. Comparing the PCMs for the two samples, we see that the amplitude of the warp for the RGB stars is larger than for the OB stars at a given radius, in the sense that the centroid of the peak that traces the warp (cross) reaches co-latitudes close to $4\degr$ for the RGBs compared to $\sim2\fdg5$ for the OBs, at $r_\gal\sim14$~kpc. 

\begin{figure}
\begin{center}
 \includegraphics[width=\columnwidth]{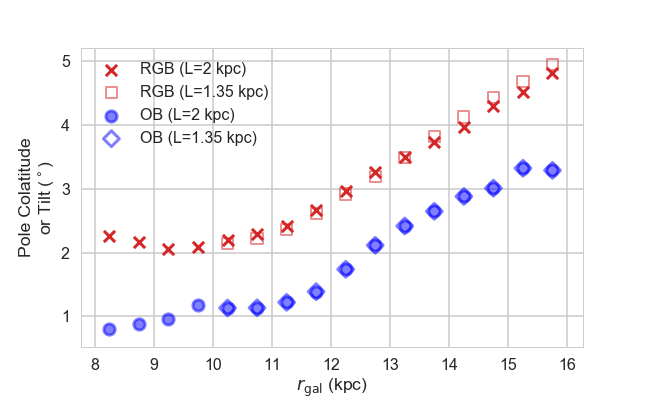}
\caption{Comparison of results for the \ngc~method obtained for two different prior scale-lengths, L=$2$ kpc (crosses and filled circles) and L=$1.35$ kpc (open squares and open diamonds) in red for the RGB sample and in blue for the OB sample. We show the inclination angles of the main \ngc~PCM peak (crosses in Fig.~\ref{fig:PCM_data}) as a function of $r_\gal$. The plateau in the tilt angle observed for the RGBs ($r_\gal\lesssim 9$~kpc) is  due to contamination from the central peak affecting the centroid when the two peaks are close.} 
\label{fig:OB_RGB_ngc_robust_dist}
\end{center}
\end{figure}

These tilt values estimated from the PCMs are compatible with those shown by the Y-Z density plot in Fig.~\ref{fig:projZ} at the same distance. The behaviour of the tilt angle as a function of $r_\gal$ is summarised in Fig.~\ref{fig:OB_RGB_ngc_robust_dist}, where the larger amplitude of the tilt angle for the RGB compared to the OB is evident, at all radii. The figure also compares these results with those obtained with a shorter prior scale length ($L=1.35$~kpc); for the two tracers both sets are in excellent agreement, showing our results are robust against the choice of the distance estimator.

\begin{figure}
\begin{center}
\includegraphics[width=\columnwidth]{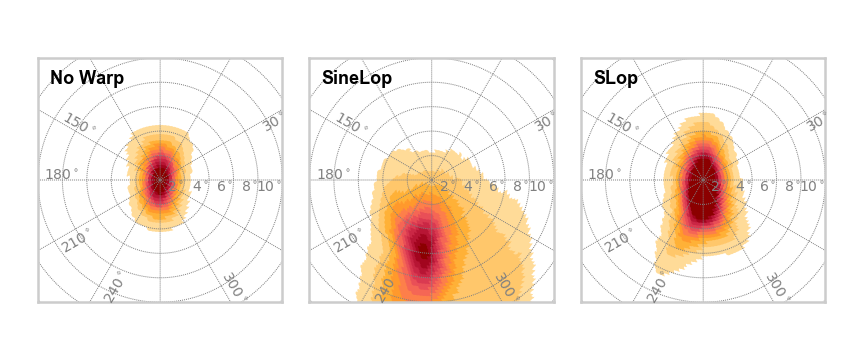}
 \caption{ \ngc~PCMs for a GaiaDR2 mock catalogue (with errors, extinction and selection function) of test particles at $r_\gal$ between 13--14~kpc, for the flat disc (\emph{left}) and the Sine Lopsided warp (\emph{right}).  Note the asymmetry in the SLop model's PCM signature, which is due to the stars lost in the I and IV quadrant when the effect of extinction and the selection function are included. The corresponding PCMs for the SineLop and SLop models without the effect of errors, extinction and selection function is shown for comparison in the Appendix in Fig.~\ref{fig:PCMs_ensemble}}. 
\label{fig:PCMs_mocks}
\end{center}
\end{figure}

Figure~\ref{fig:PCMs_mocks} shows the predictions given by the three simple warp models applied to the same mock catalogues as in Fig.~\ref{fig:Lonkinsamples_mock}, at the galactocentric ring $r_\gal=13-14$~kpc. The simultaneous signals observed at the north pole and at increasing latitude seem incompatible with the tilted-ring-like SineLop model and closer to the prediction of the SLop model. Although the SineLop model could explain the peak whose co-latitude increases with distance, it cannot account for the signal observed in the north pole and remains fixed with distance, while additionally requiring it the line of nodes to be rotated by as much as $\sim30\degr$. On the other hand, the SLop model predicts qualitatively similar signals simultaneously: the fixed signal at the pole, and the signal at increasing co-latitude that traces the increase of the warp's mean amplitude with distance. Figure~\ref{fig:PCMs_mocks} shows this prediction for the SLop model in the right panel, with signal at azimuth $240\degr$, similar to the observed peak in the PCM\footnote{For more details, see also the theoretical PCM prediction (and its discussion) for the different warp models provided in  Appendix~\ref{s:ngc_predictions} and Figs.~\ref{fig:PCMs_circular}, \ref{fig:PCMs_ensemble} and \ref{fig:PCMs_particles}.}. This suggests the SLop model can do a better job at explaining the data, although we do caution that the details of the observed signal do not match exactly the model's predictions, possibly due to a combination of model mismatch and differences in the extinction and selection function of the real and mock catalogues. 

Also, the inclined shaped of the peak that traces the warp (centroid indicated with a cross) seems most alike the PCM signature shown for mock catalogue PCMs in the SLop model. We do stress that the theoretical predictions given by the simple warp models do not fully reproduce the features in the observed PCMs, though the S Lopsided model is the one that qualitatively better matches the data, as already the LonKin method points to.

 We emphasise that for the SLop model the mean tilt angle derived from the PCM is not trivially related to the up/down tilt angles of the model due to the overlap of the two signatures (see discussion in appendix~\ref{app:predictions}). Deriving values for these parameters would require detailed modelling. Looking at the change of slope of the tilt angle as a function of $r_\gal$ in Fig.~\ref{fig:OB_RGB_lonkin_robust_dist} we estimate the starting radius of the warp to be around $12-12.5$~kpc for the OB sample and around $10.5-11$~kpc for the RGB sample, in agreement with the reported values using the LonKin method.  

\section{The richness and complexity of the vertical motion}
\label{sec:discussion}

From Sects.~\ref{sec:spatial} and \ref{sec:kine}, we clearly confirm the presence of the Galactic warp in GaiaDR2 data. The projections of the positions to the plane perpendicular to the Sun--anti-centre line (Fig.~\ref{fig:projZ}) of the OB and RGB samples reveal that the amplitude of the warp is different according to the age of the population, being the warp in the RGB sample more prominent than in the OB sample.

The simple theoretical predictions we have shown are intended to shed some light and help interpreting the results in terms of different warp models. The two dimensional projection into the Galactic plane suggests that the  line-of-nodes in the RGB sample is shifted towards galactocentric azimuth $\theta>180\degr$ and twisted in this direction from $R>12$~kpc, though this shift could be partially induced by the extinction. The results of applying both the LonKin and \ngc~PCM methods, together with the kinematic two dimensional projections, suggest that we can indeed rule out that the Galactic warp is symmetric. Both the SineLop and SLop models predict that \emph{when the warp is lopsided, the maximum proper motion does not coincide with the line-of-nodes}, as we have clearly observed with the LonKin method and in the kinematic 2D maps, with both tracers. \citet{Poggio2018}, using similar tracers but with smaller radial and magnitude coverage, and \citet{Chen2019}, using Cepheids, note that the maximum vertical velocity they observe in their maps is not along the Sun--anti-centre line and say this might be due to the Sun not being in the line-of-nodes. Here, we trace that kinematic signature further out in the disc and give an alternative interpretation based on our model predictions: that the offset between the line-of-nodes and the line of maximum vertical proper motion is due to the lopsidedness of the disc's warp.

\begin{figure}
\begin{center}
\includegraphics[width=\columnwidth]{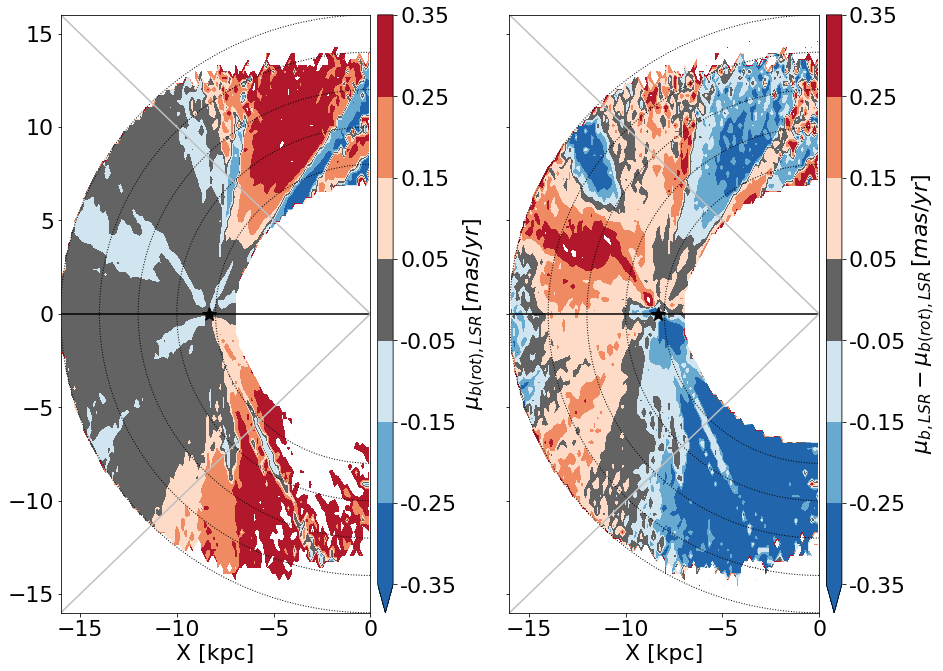}
 \caption{Contribution of a flat rotation curve (with our sample's  selection function) to the proper motion in latitude. Left panel: Given the spatial distribution of the RGB sample we compute the contribution to the proper motion in latitude with respect to the LSR of a flat rotation curve. Right panel: We subtract the proper motion in latitude the contribution due to a flat rotation curve. Bin size is the same as in the RGB sample in Fig.~\ref{fig:2dmovprop}. Black and grey lines and the black small star are the same as in Fig.~\ref{fig:2dmovprop}. }
\label{fig:mubrot}
\end{center}
\end{figure}

To be sure that the features we see in the kinematic maps and are not an artefact due to the contribution of the rotation curve to the proper motion in latitude, we make the following test. By assuming a constant rotation curve, with rotational velocity $V_c(R)=240$~\kms, we assign at each star in the RGB sample the velocity due to the flat rotation curve at the observed distance of each star and we compute the median proper motion in latitude due only to this effect. The result of this test is shown in the left panel of Fig.~\ref{fig:mubrot}. The offset between the line-of-nodes and the line of maximum vertical proper motion is not reproduced in this test, meaning that the rotation curve does not contribute in the conclusions we claim are due to the warp. We do not see any of the other effects we discuss below. To highlight this, in the right panel we also show the result of subtracting this contribution to the observed map of Fig.~\ref{fig:2dmovprop}. The resulting plot shows the same features as our original plot (Fig.~\ref{fig:2dmovprop}, right), confirming that they cannot be accounted for by a combined effect of the rotation curve and selection function of our sample.

Using both the LonKin and PCM~\ngc~methods we can establish a dependence on age for the starting radius of the warp, being that of the OB sample farther out than that of the RGB sample. More precisely, the starting radius for the young OB population would be around $12-12.5$~kpc and for the older RGB around $10.5-11$~kpc. This trend is in agreement with \citet{Amores17}, though they predict onset radii slightly smaller, of about $10$~kpc and $9$~kpc for the age bins $0-1$~Gyr and $3-7$~Gyr, respectively, and taking into account that the authors place the Sun at $R_{\odot}=8$~kpc. Using the OB stars of the TGAS-Hipparcos subsample, which extend up to $3$~kpc from the Sun, \citet{Poggio2017} find that the proper motions of the nearby OB stars are consistent with the signature of a kinematic warp, while those of the more distant stars (parallax <1 mas) are not. The authors also suggest that additional phenomena may cause systematic vertical motions that are masking the expected warp signal. \citet{Bobylev2010} uses Tycho-2 nearby ($0.3<d<1$~kpc) Red Clump stars to infer the parameters of the local warp from their kinematics. Already at this close distance to the Sun, the author detects the signal of the warp in the deformation tensor. \citet{Poggio2018} analysing similar young and old samples up to $G<16$~mag covering a smaller area in the disc, find that both samples show similar vertical motions, without providing any indication of the starting radius. \citet{Katz2018} show different vertical velocity maps for the Giant sample from those of the OB sample, again without venturing any starting radius, but indeed signalling differences between both tracers. \citet{Schonrich2018,Huang2018} using Gaia-TGAS and LAMOST-TGAS data, respectively, find the signature of the warp already at the solar neighbourhood, without studying different age tracers. Similarly, \citet{Kawata2018} select a stripe along the centre -- anti-centre direction using GaiaDR2 from $R=5-12$~kpc and finds a signal of the warp when plotting the vertical velocity as a function of radius.  

Furthermore, the presence of two relative maxima with different amplitude and a negative dip in proper motions in the LonKin plots for the RGB sample favour the SLop model as a more suitable description of the warp. The fact that the LonKin method favours an SLop shape is corroborated by the \ngc~PCM method. The over-density in the PCMs is not point-like, as would be expected from a symmetric warp; instead, it is elongated and rotated with respect to the meridians which, coupled, suggest lopsidedness. In addition, at all radii there is significant signal in the PCMs at the north Galactic pole, a feature characteristic of the SLop model alone, not observed in the SineLop model. The signature observed in the GaiaDR2 data PCMs resembles the most what we see in the mock catalogue with an SLop model (see right panel of Fig.~\ref{fig:PCMs_mocks}). As stated above, the simple tilted-ring-models used here do not reproduce every feature the kinematic data reveal, but they suggest that the SLop model could be an appropriate starting point to model the Galactic warp. The SLop model is designed such that the disc is flat not just along the line-of-nodes, like in the SS and SineLop models, but also in a region around it, out to the warp starting radius roughly (see Fig.~\ref{fig:models} for a schematic representation of the three warp models).

Figures~\ref{fig:2dmovprop}, \ref{fig:Lonkinsamples} and \ref{fig:PCM_data} clearly show the high complexity of the Galactic disc structure and it is clear that, in addition to it being detected in density in Fig.~\ref{fig:projZ}, the kinematics are in agreement with the presence of a Galactic warp. The large amount of the data and its high accuracy reveal that the kinematic maps (Fig.~\ref{fig:2dmovprop}) are not dominated by a single phenomenon, as we might expect from a Galaxy that has not reached yet equilibrium \citep{Poggio2017,Antoja2018}. The two dimensional projection of the median proper motion in latitude corrected from the Sun's peculiar motion, $\mu_{b,LSR}$, reveals complex and non-uniform trends. The radial heliocentric features are clearly related to extinction effects. But the data also highlights wave-like trends that remind those of a bending mode, apart from that given by the Galactic warp.
The bending mode, as defined recently by \citet{Widrow2014, Katz2018}, is the mean of the vertical component of the velocity measured at two parallel layers above and below the Galactic plane. In this sense, the projection of median proper motion in latitude can be viewed as a measure of the bending mode at (large) heliocentric distances where the line-of-sight velocity does not dominate the motion. Thus, from the analysis of the proper motion in latitude we can glimpse the effect of the Galactic warp (via the LonKin and \ngc~PCM methods) and the bending modes (via the two dimensional projection) at large distances. The features with median negative proper motion alternated with median positive values in the two dimensional map clearly have a wave-like pattern reminiscent of a bending mode \citep[as previously found by][]{Schonrich2018}.

Throughout all this analysis we have detected two significant features in the structure and kinematic characterisation of the RGB sample: first, an over-density of RGB stars above the Galactic plane or, possibly, an under-density of RGB stars below the Galactic plane, towards galactocentric azimuths $135\degr < \theta < 180\degr$. Second, in a similar but more constrained azimuthal range and in the outer disc, $R>12$~kpc, the so-called "blob", a minimum in proper motion in latitude, which is not predicted by the theoretical models and it is not consistent with a wave-like pattern either. We wonder whether these two feature are related. This sudden drop in proper motion is located near the region where there is also a lack of stars in the northern hemisphere at the same longitude range (see right panel of Fig.~\ref{fig:medZ}). There are not enough stars in the OB sample in that region to draw this conclusion. A similar gap in this quadrant is also seen in the open cluster distribution in the disc reported by \citet[][private communication]{CantatGaudin2018}, while \citet{Bisht2016} find a bend of the Galactic disc towards the southern latitude using young star clusters with an heliocentric distance larger than $2$~kpc in the longitude range $l = 130\degr-180\degr$. In this same longitude range and below the galactic plane, there has been long debated the origin of the TriAnd overdensity \citep{Majewski2004,Deason2014} which recently using the chemical composition has been related to the Galactic disc, extending well beyond $20$~kpc from the Galactic centre. This is a much more distant feature than the "blob", but our lack of data in the range from 15 to 20~kpc prevents us from concluding whether there is a connection or not between the two. From all these works, it seems that this region shows different features. So we suggest that the "blob" we detect as an overdensity below the Galactic plane with negative median proper motion, is a physical feature in the Galactic disc.

\section{Conclusions}
\label{sec:conclusion}

We have used two different tracers from GaiaDR2 data, namely a young bright population, the OB sample (1.8 Million sources), and the Red Giant Branch, the RGB sample (18 Million sources), in order to study their kinematics and relate them to the Galactic warp structure. To achieve this goal, we have used  different and complementary analyses that provide structural  (Sect.~\ref{sec:spatial}) and kinematic (Sect.~\ref{sec:kine}) information of the warp. Our results using spatial only data are the following:

\begin{enumerate}
    \item The RGB sample presents a clear asymmetry in stellar density with respect to the centre--anti-centre line at galactocentric azimuths $135\degr<\theta<180\degr$ clearly visible in both hemispheres, and not expected in an axisymmetric Galactic potential.
    \item The RGB sample shows a null median-Z stripe, the line-of-nodes, rotated with respect to the Sun--anti-centre direction towards $\theta\sim 180-200\degr$ and twisted from $R>10$~kpc. This suggests that the line-of-nodes of the Galactic warp is twisted.
    \item The GaiaDR2 data definitively confirms that the altitude of the warp increases with the age of the population. The altitude of the Galactic warp at $R=14$~kpc is of the order of $0.2$~kpc and $1.0$~kpc for the OB and RGB samples, respectively.
    \item The RGB sample reveals a slightly lopsided Galactic warp in the sense $|Z(down)|- |Z(up)|\sim 250$~pc. No solid conclusion can be established from the OB sample, for which lopsidedness seems to be much smaller.
\end{enumerate}

Combining spatial and kinematic information, we conclude:
\begin{enumerate}
    \item The median vertical proper motion, $\mu_{b,LSR}$, values towards the anti-centre show a clear vertical modulation. Whereas the OB sample has a transition from negative to positive values, the reverse is observed in the RGB sample. This is a clear first signature of the age dependency of the bending vertical motion. 
    \item The offset between the line-of-nodes ($\theta\sim 180-200\degr$) and the line of maximum vertical proper motion, detected in a significant stripe towards galactic azimuths $\theta\sim 160-170\degr$, can be naturally associated by our models to the lopsidedness of the disc's warp. 
    \item The fact that the vertical proper motion for the RGB data has a double-peak tendency in the LonKin and that it becomes negative towards the anti-centre at large galactocentric distances ($14<R<15$~kpc) together with the signature found in the PCM, are again clear indications of lopsidedness and suggests that an S Lopsided model would be a right choice as a starting model.
    \item Both the LonKin and the \ngc~PCM methods also allow us to estimate the initial radius of the warp. We find that the initial radius for the OB sample is $12<R<13$~kpc, while for the RGB sample is $10<R<11$~kpc, again with a clear age dependency. 
    \item The data also highlights wave-like trends that remind those of bending and breathing modes.
\end{enumerate}

The kinematic maps of both samples reveal a complex disc, not in equilibrium, in agreement with recent works \citep[e.g.][]{Antoja2018}, that reflect multiple phenomena. New models with a higher degree of complexity and methods, flexible enough to be able to capture the complexity of the disc kinematics, are required.

\section*{Acknowledgements}
The authors thank the anonymous referee for a thorough review that helped improve the manuscript.
The authors wish to thank E. Masana and J.M. Carrasco their help in the computation of $A_G$ for the full GaiaDR2 sample. This work has made use of data from the European Space Agency (ESA) mission
{\it Gaia} (\url{https://www.cosmos.esa.int/gaia}), processed by the {\it Gaia}
Data Processing and Analysis Consortium (DPAC,
\url{https://www.cosmos.esa.int/web/gaia/dpac/consortium}). Funding for the DPAC
has been provided by national institutions, in particular the institutions
participating in the {\it Gaia} Multilateral Agreement.
This work was supported by the MINECO (Spanish Ministry of Economy) - FEDER through grant
ESP2014-55996-C2-1-R and MDM-2014-0369 of ICCUB (Unidad de Excelencia "Mar\'{i}a de Maeztu"), the European Community's Seventh Framework Programme (FP7/2007-2013) under grant agreement GENIUS FP7 - 606740. We also acknowledge the team of engineers (GaiaUB-ICCUB) in charge to set up and maintain the Big Data platform (GDAF) at University of Barcelona. CM is grateful for the hospitality and support from ICCUB and IA-UNAM, where part of this research was carried out. We thank the PAPIIT program of DGAPA/UNAM for their support through grant IG100319. The authors acknowledge the use of TOPCAT \citep{Taylor2005} throughout the course of this investigation.




\bibliographystyle{aa}
\bibliography{mywarpbib} 



\begin{appendix}
\section{Details on the sample selection}
\label{app:sample}
\subsection{Cut in $f_{obs}$}
\label{sec:app_dist}
\begin{figure}
\begin{center}
 \includegraphics[width=\columnwidth]{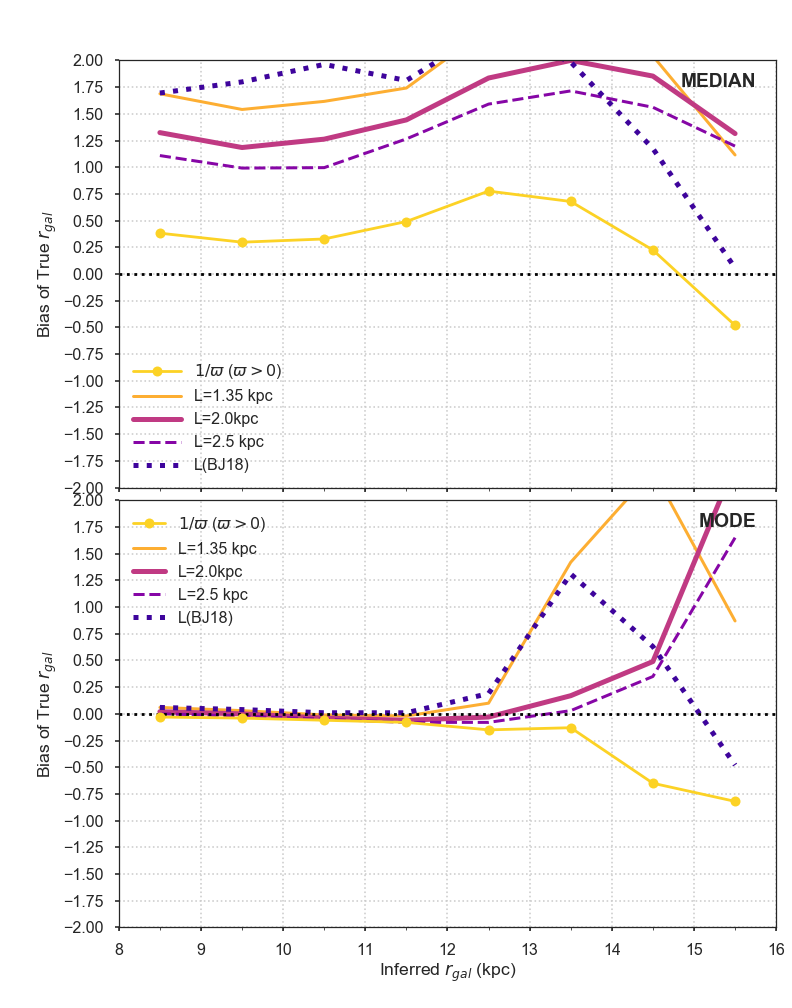}
 \caption{Same as Fig.~\ref{f:RtruDist_L_medmod_DR2} without any cut in $f_{obs}$. }
\label{fig:RtruDist_nocut}
\end{center}
\end{figure}

Here we show how the proposed cut in relative error in parallax, namely $f_{obs}$, is appropriate in order to have a distance estimator with a minimum bias in distance at large galactocentric distances. In Fig.~\ref{fig:RtruDist_nocut} we show the same as in Fig.~\ref{f:RtruDist_L_medmod_DR2} without performing any cut on $f_{obs}$. The bias in distance at large galactocentric distances is larger when using any of the scale-lengths of the exponentially decreasing prior, either using the median or the mode. 

Indeed, without performing any cut in $f_{obs}$, the number of OB and RGB stars increases up to 3948339 and 32506963, respectively, which corresponds to an increase of $47\%$ and $55\%$, respectively. However, the bias in distance obtained at large galactocentric distances is larger. We, therefore, decide to keep the proposed cut at $f_{obs}<0.5$.

\subsection{The observed HR diagram}
 \label{app:app_HR}
\begin{figure}
\begin{center}
 \includegraphics[width=\columnwidth]{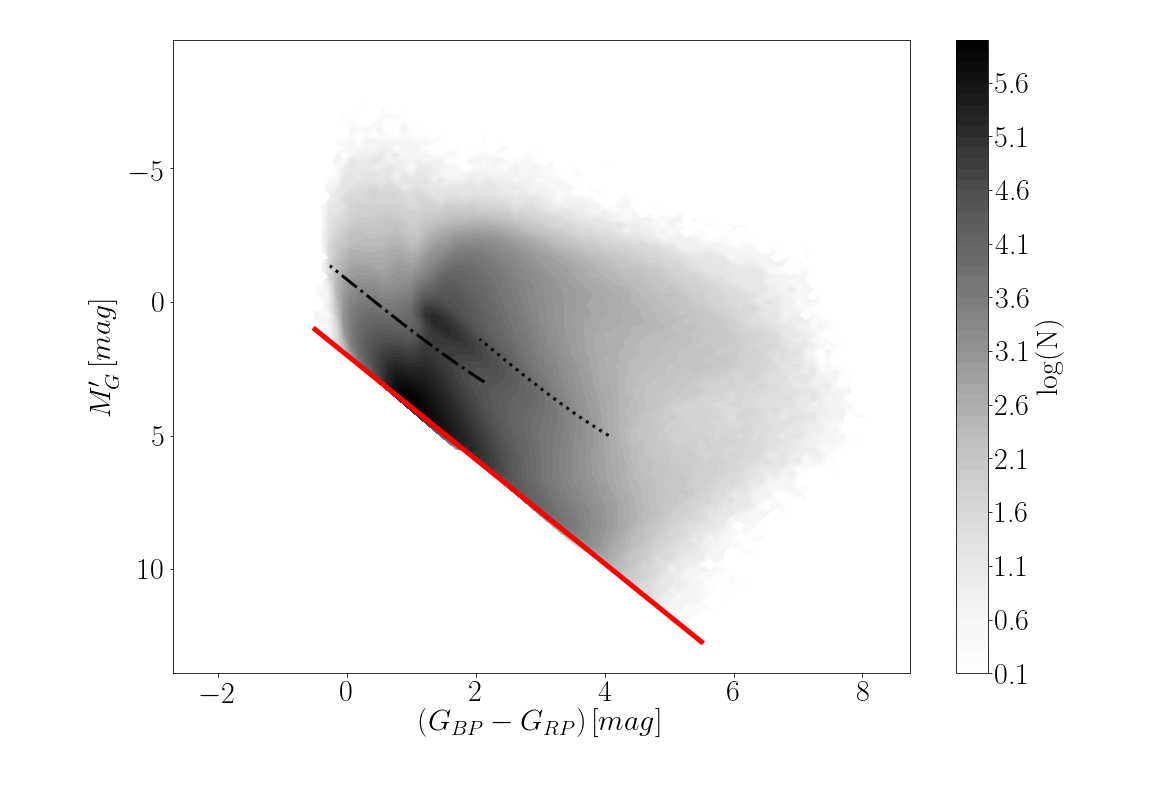}
 \caption{Observed HR diagram not corrected for absorption or extinction: $M_G^{\prime}$ as a function of the observed colour. The red line shows the cut in $M_G^{\prime}$ and the two black dotted and dot-dashed lines show two examples of the extinction line of $5$mag. }
\label{fig:HRobs}
\end{center}
\end{figure}

We describe here in detail all the steps we have performed until we obtained the de-reddened Hertzsprung-Russel diagram (HR diagram) from which we have defined the All-Stars. As mentioned in the main text, first we select from GaiaDR2 all stars up to magnitude G=20, with an available parallax measurement and with $f_{obs}<0.5$. This first cut reduces the GaiaDR2 sample to $383,510,799$ sources. For these stars, we compute the distance using a Bayesian estimator (see Sect.~\ref{sec:distance_estimators}) and $M_G^{\prime}=G-5\log10(d)+5$, the absolute G magnitude of the star uncorrected for absorption. We now want to remove cool main sequence stars from the sample. To do so, we infer what would be the extinction line of GaiaDR2 data, from the relation between the absorption in G, $A_G$, and the reddening, E($\bp$-$\rp$) from \citet{Arenou2018}, which is about $1.95$ and it will determine the slope of the extinction line. In Fig.~\ref{fig:HRobs} we show the observed HR diagram, $M_G^{\prime}$ as a function of the observed colour ($\bp$-$\rp$), and in black dotted lines two extinction curves for $5$~mag of absorption. So a hypothetical star with $M_G^{\prime}=5$, observed colour ($\bp$-$\rp$)$=4$ and absorption $A_G\le 5$ would have an extinction curve like the black dotted one. Therefore, we perform a cut parallel to the extinction line $M_G^{\prime}$ < 1.95*($\bp$-$\rp$)+2., and with a zero point chosen so that a typical star with observed colour ($\bp$-$\rp$)=0. would have $M_G^{\prime}=2$ to include all the OB sample of the main sequence. This cut is shown as a red curve in Fig.~\ref{fig:HRobs}. This second cut reduces the sample to $86,814,618$ sources. From this sample we remove the stars in the Large and Small Magellanic Clouds, the Andromeda Galaxy (M31), and five globular clusters visible in the (l,b) projected density: NGC~6205, NGC~6341, NGC~7078, NGC~1851, NGC~7089 \citep{Helmi2018}, which are distant objects in the halo that may have large uncertainties in parallax that can create artificial over-densities in the vertical space.

From the All-Stars, we select the OB and RGB using the same recipe as in \citet{Katz2018} (see Eqns.~\ref{eq:selectionOB} and \ref{eq:selectionRGB}). In order to assess the possible contamination in each sample, we use the Gaia mock catalogue of GUMS \citep{Robin12} with the errors in parallax scaled to GaiaDR2 performances as in \citet{Luri2018}. Since GUMS contains the spectral type, we are able to assess the possible contamination our strategy may include. Therefore, we perform the same cuts described above and we select the OB and RGB samples again as in Eqns.~\ref{eq:selectionOB} and \ref{eq:selectionRGB}. We estimate that $91\%$ of the stars in the OB sample are of the spectral type O, B and A0, so only a $9\%$ of the OB sample are contaminants. The RGB sample includes $92\%$ of sub-giants, giants and super-giants, so only $8\%$ of the stars in the RGB sample are contamination, entirely due to dwarf stars.

\section{Warp models}
\label{app:models}
We describe the three warp models we use in this study to compare with data: one symmetric as in \citet{Abedietal14}, the "Simple Symmetric warp", and two lopsided models first used in this work: the "Sine Lopsided" and the "S Lopsided" models. In Fig.~\ref{fig:models} (first to third row), we show a schematic representation in 3D (left column) and in projection along the line-of-nodes (right column) of the three warp models and an example of the twisted S Lopsided model (fourth row), whose mathematical forms are described in detail in the following subsections.

\subsection{The Simple Symmetric warp}\label{s:ss_model}
The Simple Symmetric warp (SS) is already described in detail in \citet{Abedietal14}. It consists of splitting the Galactic disc into galactocentric rings and tilting each one by an angle $\psi$ as a function of the galactocentric radius. The expression for the tilt angle is:

\begin{equation}
\psi_{SS}(R;R_1,R_2,\alpha,\psi_2) = \psi_2\,f(R;R_1,R_2,\alpha)
\label{eq:SS}
\end{equation}

where the function $f(R)$ denotes the dependence of the tilt angle with radius and we express it as a power law:

\begin{equation}
f(R;R_1,R_2,\alpha) = \begin{cases}
0, & R\le R_1\\ \left({{R-R_1}\over{R_2-R_1}}\right)^\alpha, & R_1<R<R_2\\ 1, & R\ge R_2
\label{eq:fR}
\end{cases}
\end{equation}
where $R_1$ and $R_2$ are the initial and final radius of the warp, and the exponent $\alpha$ fixes the shape of the power law.

We notice that once the warp parameters $R_1$, $R_2$, and $\alpha$ have been fixed, the tilt angle only depends on the galactocentric radius $R$, and has a constant maximum amplitude at $R \ge R_2$ of $\psi_2$. This defines a symmetric warp with respect to the line-of-nodes, i.e. the maximum amplitude of the warp is the same at corresponding points on both sides of the line-of-nodes. According to observational evidence, the up (down) side of the Galactic warp is towards the northern (southern) hemisphere \citep[e.g.][]{Oort1958,LopezCorredoira2002,Levine2006}.

\subsection{The Sine Lopsided Model}\label{s:sine_model}

\begin{figure}
\begin{center}
 \includegraphics[width=\columnwidth]{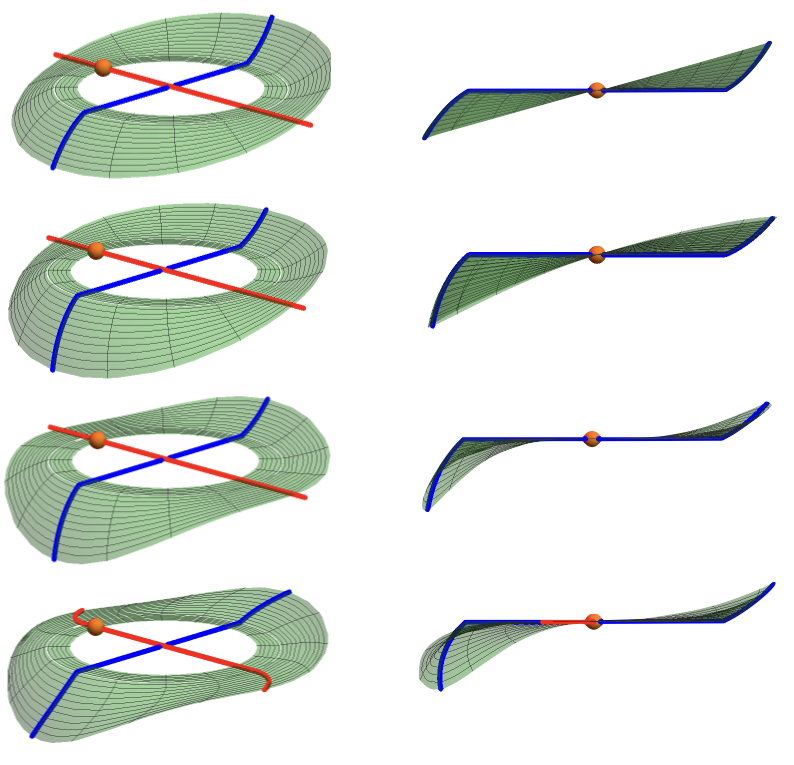}
 \caption{Schematic plot of a warped disc according to the three warp models (see text): First row: Simple Symmetric warp. Second row: Sine Lopsided warp. Third row: S Lopsided warp. Fourth row: S Lopsided warp with a twisted line-of-nodes. The red line shows the line-of-nodes and the orange sphere marks the position of the Sun. The blue line, perpendicular to the line-of-nodes, shows the maximum amplitude of the warp. The amplitude values for the schematic plot have been increased for the sake of clarity. }
\label{fig:models}
\end{center}
\end{figure}

The Sine Lopsided model (SineLop) is a variation of the Simple Symmetric model. It introduces a dependence on the galactocentric azimuth, $\theta$, while keeping the same radial dependence as the Simple Symmetric warp model. Therefore, the maximum amplitude of the warp changes on either side of the line-of-nodes.  The tilt angle is given by: 
\begin{equation}
\psi_{SL}(R, \theta;R_1,R_2,\alpha,\psi_{up},\psi_{down}) = 
f(R;R_1,R_2,\alpha)\,  [A+B\sin(\theta)]
\label{eq:SL}
\end{equation}
where $A=(\psi_{up}+\psi_{down})/2$, $B=(\psi_{up}-\psi_{down})/2$ and $\psi_{up}$ and $\psi_{down}$ are the maximum amplitudes of the warp at the up and down warps, respectively, at the azimuth perpendicular to the line-of-nodes. Note that, using this expression, rings are not planar any more, i.e. a warped ring when seen edge-on is not a flat line, but it is slightly bent at the line-of-nodes, as illustrated by the edge-on view shown in Fig.~\ref{fig:models} (second row, right). This curvature is caused by the sine dependence on the azimuth, that allows us to introduce an arbitrary up/down maximum amplitude asymmetry in a smooth (continuous and differentiable) manner across the line of nodes, which is not attainable by just setting different tilt angles at either side of the line of nodes in the tilted ring model of the previous section. Hence, the expression of Eq.~\ref{eq:SL} provides a lopsided disc (see second row panels of Fig.~\ref{fig:models}). In the Galactic disc's warp, the amplitude in the up side is larger than in the down side \citep{Levine2006,Marshall06}.

\subsection{The S Lopsided Model}\label{s:slop_model}

The S lopsided model (SLop) is designed to have an S-shape when seen edge-on, that is, it is flat at the line-of-nodes. It is a variation of the Sine Lopsided model with a different dependence on the galactocentric azimuth. We first define an angle function given by:
\begin{equation}
g(\theta) = \begin{cases}
\psi_{up}, & 0\le\theta < 180^\circ\\ \psi_{down}, & 180^\circ\le\theta < 360^\circ
\end{cases}
\label{eq:gtheta}
\end{equation}
where, as previously, $\psi_{up}$ and $\psi_{down}$ are two constant angles which are the maximum amplitudes of the warp at the up and down sides, respectively.

The tilt angle for this model is given by:
\begin{equation}
\psi_{SLop}(R, \theta;R_1,R_2,\alpha,\psi_{up},\psi_{down}) = f(R;R_1,R_2,\alpha) g(\theta)\sin^2(\theta)
\label{eq:SLop}
\end{equation}

\subsection{Twisted versions}
In all warp models, the line-of-nodes can either be a straight line aligned with the Sun -- Galactic centre line or it can be twisted an angle $\phi$:

\begin{equation}
\phi(R;R_1,R_2,\gamma,\phi_{max}) = \begin{cases}
0, & R\le R_1\\ \phi_{max} \left({{R-R_1}\over{R_2-R_1}}\right)^\gamma, & R_1<R<R_2\\ \phi_{max}, & R\ge R_2
\label{eq:phiR}
\end{cases}
\end{equation}

where it increases with radius from the x-axis (see last row of Fig.~\ref{fig:models}. 

\section{Warp transformations}
\label{app:veltrans}

The purpose of this section is to develop the equations that provide the spatial and velocity coordinates for particles in a warped disc. We apply a warp transformation to the spatial coordinates of particles in a flat disc, and then find the corresponding transformation for the velocities. As we will see, velocities require a different transformation.

\subsection{The spatial transformation}
\begin{figure}
\begin{center}
 \includegraphics[width=\columnwidth]{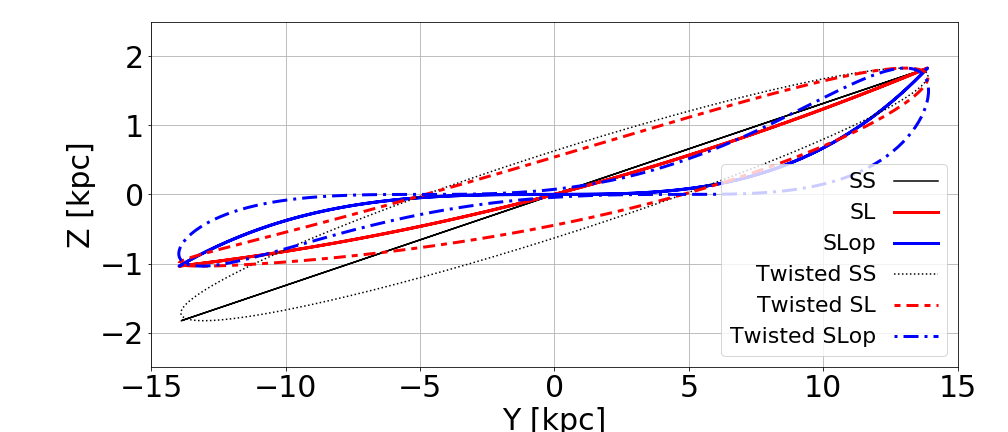}
 \caption{Y-Z projection of a tilted circular orbit at R=14kpc using each of the three warp models: simple warp (red solid line), simple lopsided warp (blue short dashed line) and S lopsided warp (green long dashed line). }
\label{fig:orbits}
\end{center}
\end{figure}

As mentioned in the previous section, to warp a flat disc we use a rotation by an angle $\psi$ with respect to the positive $x$--axis (see Fig.~\ref{fig:orbits}):
\begin{equation}
\begin{pmatrix}x'\\ y'\\ z'\end{pmatrix}=\begin{pmatrix}
1 & 0 & 0\\ 0& \cos(\psi) & -\sin(\psi)\\ 0& \sin(\psi)&\cos(\psi)\end{pmatrix}
\begin{pmatrix}x\\ y\\ z\end{pmatrix}=
\begin{pmatrix}x\\ y\cos(\psi)-z\sin(\psi)\\ z\cos(\psi)+y\sin(\psi)\end{pmatrix}
\label{eq:sptransf}
\end{equation}
Where $(x',y',z')$ are the warped coordinates and $(x,y,z)$ are the Cartesian coordinates in the original flat disc. So the $x$--axis will be the line of nodes (our warp does not include twisting, which implies a non--straight line of nodes).

In the case of a symmetric warp, the tilt angle $\psi$ is a function of galactocentric radius only, thus generating a family of concentric, flat, tilted rings. In the general case of a lopsided warp, the tilt angle will be also a function of the azimuthal, galactocentric angle $\theta$ (measured with respect to the positive $x$--axis), thus introducing the bending needed to produce the asymmetric warp. In the next subsection we work with the latter, more general case.

\subsection{The velocity transformation}
One may think that velocities are transformed using the same transformation used for positions, but this is not the case. Velocities are spatial displacements per time, and so they are transformed as:

\begin{equation}
v'_i={{dx'_i}\over{dt}}= {{\partial x'_i}\over{\partial x_j}} {{dx_j}\over{dt}} ={\mathbf \Lambda'}_{i,j}v_j
\label{eq:TempVelCoordTransf}
\end{equation}
where summation over repeated indices is implied.

In order to compute the  ${\mathbf\Lambda}'$ matrix, first we obtain the derivatives of the tilt angle in Cartesian coordinates. We note that the dependence of the tilt angle is given by:

\begin{equation}
\psi=\psi(R,\theta),\quad\text{with}\,\,\, R=(x^2+y^2)^{1/2},\,\,\,  \theta=\arctan{y/x}
\end{equation}

The components of the Jacobian of the transformation between polar and Cartesian coordinates are:
\begin{equation}
 {{\partial R}\over{\partial x}}={{x}\over{R}},\quad {{\partial R}\over{\partial y}}={{y}\over{R}},\qquad
{{\partial \theta}\over{\partial x}}=-{{y}\over{R^2}},\quad {{\partial \theta}\over{\partial y}}={{x}\over{R}}
\end{equation}

The partial cartesian derivatives of the tilt angle are then:
\begin{equation}
{\partial\psi\over{\partial x}} =\dpsidR {{\partial R}\over{\partial x}} + \dpsidth {{\partial\theta}\over{\partial x}} = {x\over{R}}\dpsidR -{y\over{R^2}}\dpsidth \\
\label{eq:derivpsix}
\end{equation}
\begin{equation}
{\partial\psi\over{\partial y}} =\dpsidR {{\partial R}\over{\partial y}} + \dpsidth {{\partial\theta}\over{\partial y}} = {y\over{R}}\dpsidR +{x\over{R^2}}\dpsidth
\label{eq:derivpsiy}
\end{equation}

With these, we can now proceed to get the individual components of the ${\mathbf\Lambda}'$ matrix using equation~(\ref{eq:sptransf}):
\begin{equation*}
{\mathbf\Lambda'}_{1,1}=(\partial x'/\partial x)=1,
\end{equation*}
\vskip-12pt
\begin{equation*}
{\mathbf\Lambda'}_{1,2}=(\partial x'/\partial y)=0,
\end{equation*}
\vskip-12pt
\begin{equation*}
{\mathbf\Lambda'}_{1,3}=(\partial x'/\partial z)=0
\end{equation*}
\vskip-12pt
\begin{equation*}
{\mathbf\Lambda'}_{2,1}={\partial \over{\partial x}}[y\cpsi-z\spsi]=-{\partial\psi\over{\partial x}}[z\cpsi+y\spsi]
\end{equation*}
\vskip-12pt
\begin{equation*}
{\mathbf\Lambda'}_{2,2}={\partial \over{\partial y}}[y\cpsi-z\spsi] =\cpsi-{\partial\psi\over{\partial y}}[z\cpsi+y\spsi]\\
\end{equation*}
\vskip-12pt
\begin{equation*}
{\mathbf\Lambda'}_{2,3}={\partial \over{\partial z}}[y\cpsi-z\spsi] =-\spsi
\end{equation*}
\vskip-12pt
\begin{equation*}
{\mathbf\Lambda'}_{3,1}={\partial \over{\partial x}}[z\cpsi+y\spsi]={\partial\psi\over{\partial x}}[y\cpsi-z\spsi]
\end{equation*}
\vskip-12pt
\begin{equation*}
{\mathbf\Lambda'}_{3,2}={\partial \over{\partial y}}[z\cpsi+y\spsi] =\spsi+{\partial\psi\over{\partial y}}[y\cpsi-z\spsi]
\end{equation*}
\vskip-12pt
\begin{equation*}
{\mathbf\Lambda'}_{3,3}={\partial \over{\partial z}}[z\cpsi+y\spsi] =\cpsi
\end{equation*}

If we now define the following functions: $P\equiv z\cpsi+y \spsi,\quad Q\equiv y\cpsi-z\spsi$, the ${\mathbf\Lambda}'$ matrix can be written in a very compact way as:
\begin{equation}
{\mathbf\Lambda}'=\begin{pmatrix}1&0&0\\ -P\dpsidxl & \cpsi-P\dpsidyl & -\spsi\\ Q\dpsidxl & \spsi+Q\dpsidyl & \cpsi\end{pmatrix}
\label{eq:Lambdafinal}
\end{equation}
where the partial derivatives of $\psi$ are given by equations~(\ref{eq:derivpsix}) and (\ref{eq:derivpsiy}).

Note that in the case of a fixed tilt angle (inclined plane), or if we consider pure circular rotation along flat, tilted rings, the partial derivatives are null and the ${\mathbf\Lambda}'$ matrix coincides with the spatial transformation matrix of equation~(\ref{eq:sptransf}). But in the general case, the $\psi$-derivatives make the velocity transformation different from the spatial one.

The transformed (but unscaled) velocity components are then given by:
\begin{equation}
v'_x = v_x
\end{equation}
\vskip-12pt
\begin{equation}v'_y = -v_x P\dpsidxl +v_y[\cpsi-P\dpsidyl]-v_z\spsi
\end{equation}
\vskip-12pt
\begin{equation}
v'_z = v_x Q\dpsidxl +v_y[\spsi+Q\dpsidyl]+v_z\cpsi
\label{eq:unscalv}
\end{equation}

Now, since the velocity transformation redirects the velocity vectors, but should not change their magnitudes, we need to scale the velocity components making sure that the $v_x$ remains constant. In other words, we need to satisfy:
\begin{equation}
v_x^2+v_y^2+v_z^2=(v_x^t)^2+(v_y^t)^2+(v_z^t)^2\quad \mathrm{and}\quad v_x=v^t_x
\label{eq:restric}
\end{equation}

This implies:
\begin{equation}
v_y^2+v_z^2=(v_y^t)^2+(v_z^t)^2
\end{equation}

Which means that the scaling for the transformed velocity components is:
\begin{equation}
v_x^t = v_x,\quad v_y^t = v'_y\times v_{scale},\quad  v_z^t = v'_z\times v_{scale},
\label{eq:vscal}
\end{equation}
where the scaling factor is:
\begin{equation}
v_{scale} = \sqrt{{v_y^2+v_z^2}\over{(v'_y)^2+(v'_z)^2}}
\label{eq:vfactor}
\end{equation}

It only remains to specify $(\partial\psi/\partial R)$ and $(\partial\psi/\partial\theta)$ for each of the three warp models used.
\medskip

For the Simple Symmetric model, the tilt angle derivatives are simply:
\begin{equation}
{{\partial\psi_{SS}}\over{\partial R}}=
\psi_{max}{{d f}\over{dR}}
,\quad {{\partial\psi_{SS}}\over{\partial \theta}}=0
\label{eq:dpsiSS}
\end{equation}
where $f(R)$ is as in Eq.~\ref{eq:fR}.
\medskip

For the Sine Lopsided model, the derivatives of $\psi_{SL}$ are:
\begin{equation}
{{\partial\psi_{SL}}\over{\partial R}}={{df}\over{d R}}\,[A+B\sin(\theta)],\quad
{{\partial\psi_{SL}}\over{\partial\theta}}=f(R)\,B \cos(\theta)
\label{eq:dpsiSL}
\end{equation}
\medskip

Finally, for the S Lopsided model, the derivatives are:
\begin{equation}
{{\partial\psi_{SLop}}\over{\partial R}}={{df}\over{d R}}\,g(\theta)\sin^2(\theta),\quad
{{\partial\psi_{SLop}}\over{\partial\theta}}=f(R)g(\theta)\, \sin(2\theta)
\label{eq:dpsiSLop}
\end{equation}

where $g(\theta)$ is as in Eq.~\ref{eq:gtheta}. 

In the transformation outlined here, the twist is not included. This for reasons of brevity and clarity. To include a twist, an additional rotation around the positive z-axis by an angle $\phi$ should be added to the rotation matrix (Eq.~\ref{eq:sptransf}) according to equation (Eq.~\ref{eq:phiR}). The resulting combined rotation matrix should be used to compute the new version of the $\mathbf\Lambda’$ matrix.

\section{Prediction of the kinematic signature of the warp as seen by the Lonkin and \ngc~methods}
\label{app:predictions}

In this section, we discuss the expected theoretical signatures of the three warp models described in Sect.~\ref{app:models} for the LonKin and the \ngc~PCM methods.

To predict and understand the kinematic signatures of the different warp models we use three different test scenarios of increasing complexity: a circular orbit, a random realization of warp models and a test particle simulation. 

We fix the free parameters of the warp models to the values derived from \citet{Amores17} for a population of Red Clump stars (hereafter, RC): $R_1=10.1$~kpc, $R_2=14$~kpc, $\alpha=1.1$ and $\psi_2=7\fdg5$, $\psi_{up}=7\fdg5$ and $\psi_{down}=4\fdg25$. In the case of the twisted line-of-nodes, we fix $\phi_{max}=20\degr$ and $\gamma=1.1$. In the remainder of this section we will assume these numerical values for the warp model parameters.

\subsection{The three test scenarios}\label{s:test_scenarios}

The first scenario is the simplest case of a planar circular orbit warped following each of the three warp models. We take a circular orbit at $R\gal=R_2=14$~kpc, with circular velocity of $V_c(R_2)=215$~\kms, assuming the rotation curve of \citet{Allen1991}. The equations we use to warp positions and velocities are detailed in appendix~\ref{app:veltrans}.

The second scenario is a random realisation, obtained by taking an initial set of particles in a planar disc and applying positions and velocities transformations to each of the particles. As the initial set, we take the set of disc Red Clump initial conditions as in \citet{Abedietal14,RomeroGomez2015}. They are a subset of the red giant branch stars, well defined in the HR diagram. They are generated according to a disc population of RC stars with the corresponding velocity dispersion\footnote{As in \citet{RomeroGomez2015}, we fix the velocity dispersion of RC K-giant stars at the Sun position to be $30.3$~\kms, $23.6$~\kms and $16.6$~\kms in the radial, tangential and vertical directions, respectively \citep[][and references therein]{BT2008}, and a constant scale-height of $300$~pc \citep{Robin86}.} and with the density in statistical equilibrium with the Miyamoto-Nagai disc potential  \citep[see appendix A of][]{RomeroGomez2015}. We note that by construction, the realisation has non-zero radial velocity dispersion.  

The third scenario considered is the test particle simulation, for which we take the same initial conditions for a disc RC population as we did for the realisation. The number of particles in the simulation is chosen to have a realistic surface number density of RC stars in the Solar neighbourhood \citep{Czekaj2014,RomeroGomez2015}. The integration strategy is the same as in \citet{Abedietal14}, where we integrated the initial conditions 
to obtain a test particle simulation in statistical equilibrium with the imposed potential. This strategy consists of integrating the particles in a flat disc potential for two periods (of an orbit at $R\gal=14$~kpc), then integrating for another five periods while the disc potential is changed towards the chosen warp model, and in the third step, integrating two further periods to allow particles to settle into the final warped disc potential. We want to stress that the success of this strategy to obtain a set of particles in statistical equilibrium depends on the warp model. We have confirmed it is successful for the Simple Symmetric and the Sine Lopsided models, but the test particle simulation with the S Lopsided model imposed does not reach statistical equilibrium, even if we increase the number of periods in which the potential is warped with time. 
In this case, we cannot ensure that the positions and velocities of the test particles using the S Lopsided model will inherit those established by the model. 

We generate a GaiaDR2 mock catalogue up to magnitude G = 20 using the prescription of the astrometric, photometric and spectroscopic formal errors for Data Release 2\footnote{A fortran code to generate the Gaia errors is provided in \texttt{https://github.com/mromerog}}. We simulate errors for a mission time of $22$ months and, for the bright stars (G<13), we include a multiplicative factor of $3.6$ to the error in parallax to match the distribution of uncertainties as a function of the G magnitude observed in GaiaDR2 data.

\subsection{Theoretical signature in the LonKin method}\label{s:lonkin_predictions}

\begin{figure}
\begin{center}
 \includegraphics[width=\columnwidth]{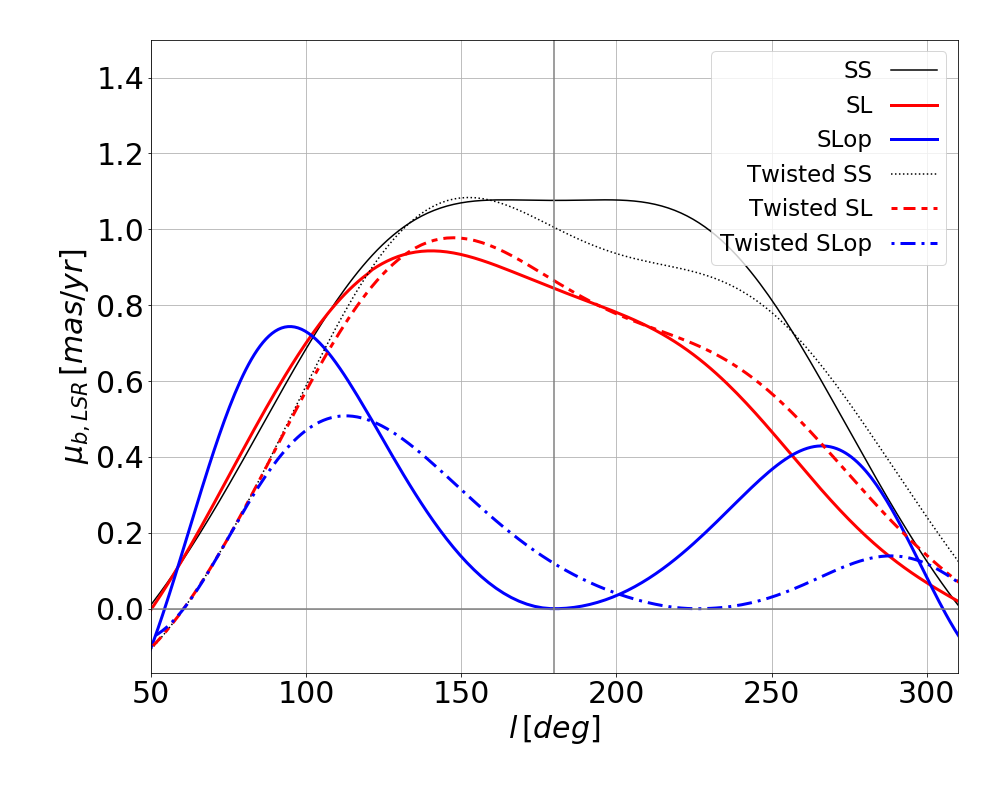}
 \caption{LonKin method applied to a warped circular orbit at $R\gal=14$~kpc using each of the three warp models: Simple Symmetric warp, SS (black solid line), the Sine Lopsided warp, SL (red solid line) and the S Lopsided warp, SLop (blue solid line). The twisted version of the three models is shown by dotted, dashed and dotted-dashed lines, respectively.}
\label{fig:Lonkinorbits}
\end{center}
\end{figure}

\begin{figure}
\begin{center}
\includegraphics[width=\columnwidth]{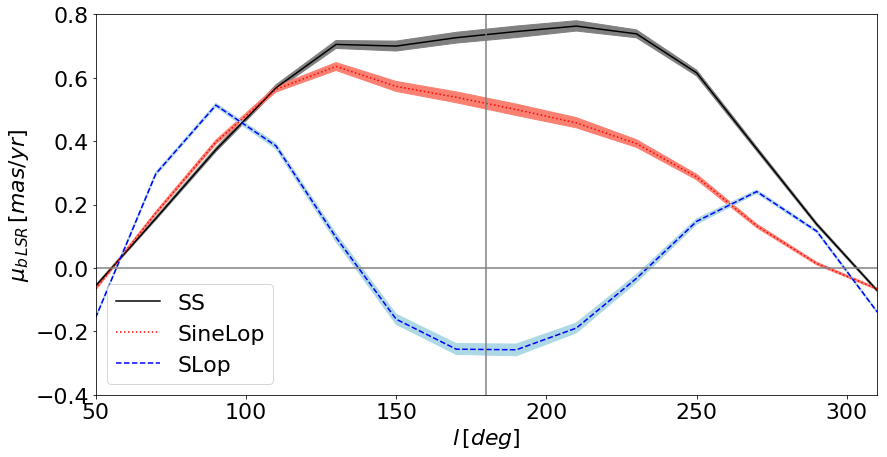}
\caption{LonKin method applied to random realisations of the reference models, at the galactocentric ring $13<R<14$ kpc. Simple symmetric (SS) in black, Sine Lopsided (SineLop) in red and S Lopsided (SLop) in blue. The line shows the median of the particles in bins of $20\degr$ and the error bars (shaded areas) show the lower and upper $1\sigma$ uncertainty in the position of the median.}
\label{fig:Lonkinsamples_theo}
\end{center}
\end{figure}

\begin{figure*}
\begin{center}
 \includegraphics[width=\columnwidth]{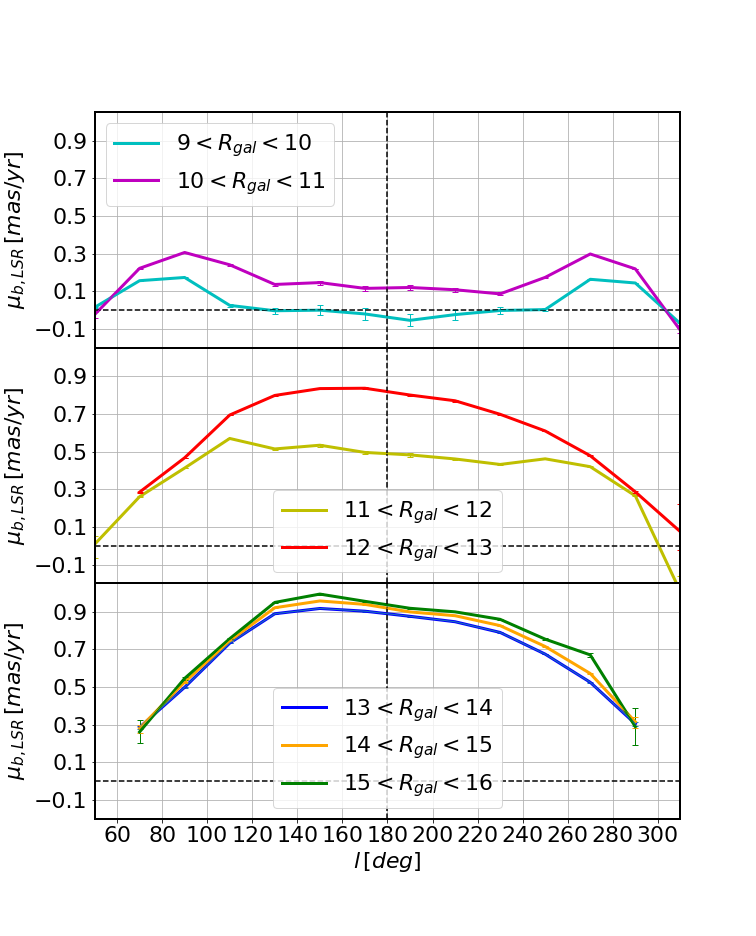}
 \includegraphics[width=\columnwidth]{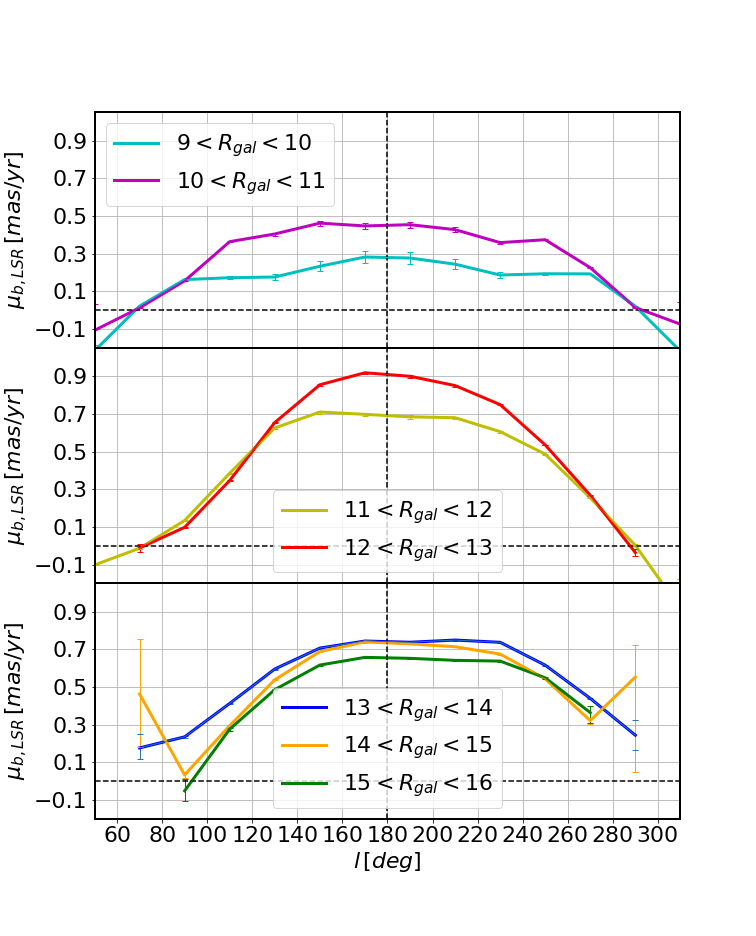}
 \caption{LonKin method applied to RC test particle simulations. Left panel: using the Sine Lopsided model. Right panel: using the S Lopsided model. From top to bottom, different radial galactocentric bins, specified in the legend, expanding from the inner to the outer disc. Horizontal dashed line shows the zero-axis while the vertical dashed line shows the anti-centre direction at $l=180\degr$. As in Fig.~\ref{fig:Lonkinsamples_theo}, the error bars show the lower and upper $1\sigma$ uncertainty of the median.}
\label{fig:LonkinRC}
\end{center}
\end{figure*}

First, we consider the simplest case of a warped circular orbit. In Fig.~\ref{fig:Lonkinorbits} we show how a warped circular orbit at $R\gal=14$kpc looks when we apply the LonKin method using the three warp  models described in Sect.~\ref{app:models}, with and without twist of the line-of-nodes. As expected from \citet{Abedi2015}, the proper motion in latitude, $\mu_{b,LSR}$, has a maximum in the anti-centre direction when the orbit is tilted using the Simple Symmetric model. As soon as the symmetry is broken, i.e. in the lopsided models, the maximum is no longer in the anti-center and the proper motion in latitude has a different behaviour depending on the model used to warp the orbit. The Sine Lopsided model simply displaces the maximum towards the longitude of the warp's maximum amplitude. When the orbit is warped using the S Lopsided model, the proper motion in latitude becomes zero at the anti-centre direction. Note that in this model, the disc is flat at the line-of-nodes. The main characteristic of the S Lopsided model is the presence of two relative maxima at $l$ around $90\degr$ and $270\degr$.

In Fig.~\ref{fig:Lonkinsamples_theo}, we apply the LonKin method to a random realisation of particles inside the cylindrical galactocentric radial bin $13-14$~kpc. As expected, the curve corresponding to the Simple Symmetric warp is essentially flat in the anti-centre direction and symmetric with respect to the line-of-nodes, while the signature of the Sine Lopsided warp displaces the maximum towards the longitudes of maximum warp amplitude and the S Lopsided warp shows two clear maxima at $l\sim 90\degr$ and $l\sim 270\degr$. We note that warping the ensemble of particles using this last model, the proper motion in latitude is no longer zero at the line-of-nodes. This fact is due to the velocity dispersion of the generated particles.

Finally, in Fig.~\ref{fig:LonkinRC} we show the result of applying the LonKin method to the test particle simulations of disc RC stars. We divide Fig.~\ref{fig:LonkinRC} in three panels according to the galactocentric radial bin of the particles. From top to bottom: $9<R<10$ and $10<R<11$ kpc; $11<R<12$ and $12<R<13$ kpc; and $13<R<14$, $14<R<15$ and $15<R<16$ kpc. The warp models used are the Sine Lopsided (left) and the S Lopsided (right). We stress here that only the test particle simulation using the Sine Lopsided model is in statistical equilibrium, while the result of the test particle simulation using the S Lopsided model will not follow the trend predicted for such model (compare with Figs.~\ref{fig:Lonkinorbits} and \ref{fig:Lonkinsamples_theo}) because it is not in statistical equilibrium.
As for the Sine Lopsided test particle simulation, we note that the median proper motion in latitude increases towards the outer disc and the outermost curve shows the expected shape for this model (compare with Figs.~\ref{fig:Lonkinorbits} and \ref{fig:Lonkinsamples_theo}), namely the maximum of the curve is shifted towards the maximum amplitude of the warp. Thus, if the disc of the Milky Way is warped following a Sine Lopsided model, we should expect a signature in the LonKin method qualitatively similar to that in the left panel of Fig.~\ref{fig:LonkinRC}. As expected, the results regarding the test particle simulation not being in statistical equilibrium using the S Lopsided model have lost the predicted signature for this model in the previous scenarios. The proper motion in latitude does not show the two relative maxima. Since the simulation is not in equilibrium, it is expected that the median values of the proper motion in latitude per longitude bins resemble those for a symmetric warp: we simply recover the increase of the proper motion with respect to the galactocentric radius (Eqs.~\ref{eq:SS} and \ref{eq:fR}), but it clearly does not show a lopsided behaviour.

\subsection{Theoretical signature in the \ngc~method}\label{s:ngc_predictions}

\begin{figure}
\begin{center}
 \includegraphics[width=0.7\columnwidth]{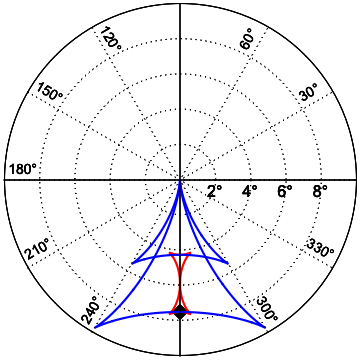}
 \caption{\ngc~PCM signatures for a circular orbit at $r_\gal=14$~kpc, for each of the three warp models: Simple Symmetric warp (black), the Sine Lopsided warp (red) and the S Lopsided warp (blue). The dotted circles correspond to parallels at co-latitudes of $2\degr, 4\degr, 6\degr, 8\degr$ and $10\degr$, with the outermost one (solid black line) corresponding to $10\degr$. }
\label{fig:PCMs_circular}
\end{center}
\end{figure}

\begin{figure}
\begin{center}
\includegraphics[width=\columnwidth]{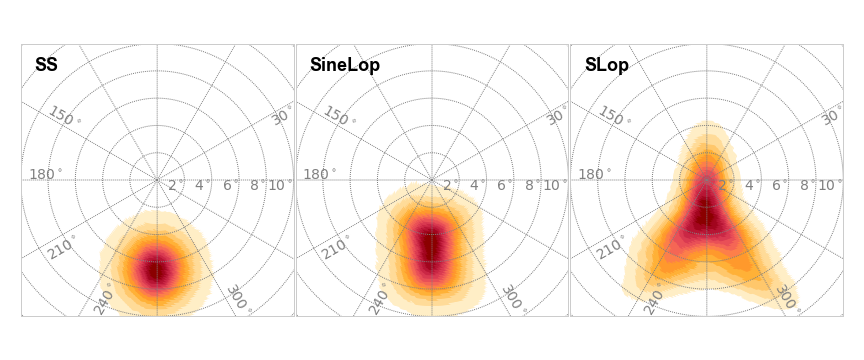} 
 \vspace{-0.5cm}
 \caption{ \ngc~PCMs for a random realisation of particles in the $r_\gal$ range 13--14~kpc, for each of the three warp models: Simple Symmetric warp (\emph{left}), the Sine Lopsided warp (\emph{centre}) and the S Lopsided warp (\emph{right}). The colour scales linearly with star counts, indicating larger values with darker colours. Dotted circles and radial lines correspond respectively to parallels at co-latitudes increasing by $2\degr$ and meridians at azimuths every $30\degr$.}
\label{fig:PCMs_ensemble}
\end{center}
\end{figure}

\begin{figure}
\begin{center}
\includegraphics[width=\columnwidth]{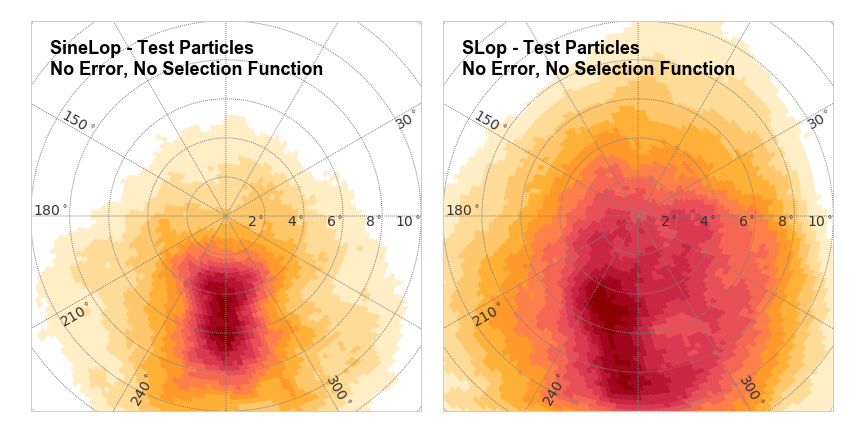}
\includegraphics[width=\columnwidth]{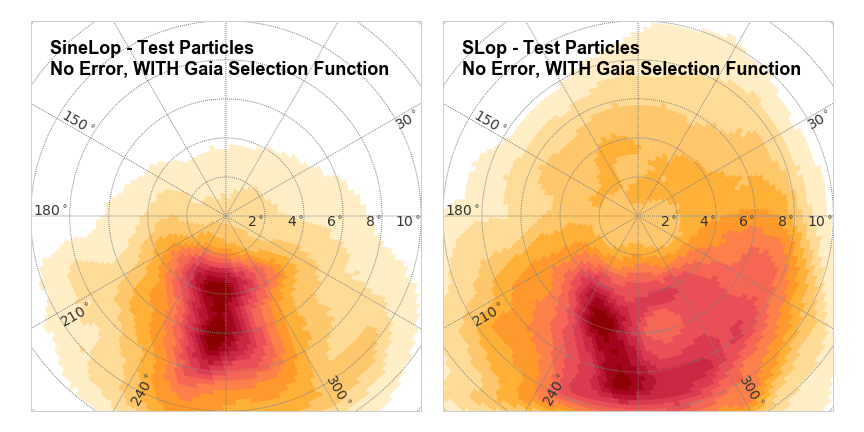}
\includegraphics[width=\columnwidth]{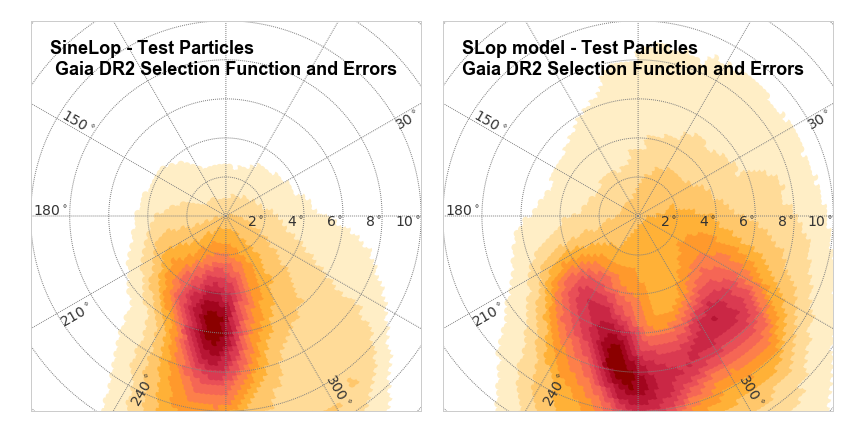} 
 \caption{\ngc~PCMs for a GaiaDR2 mock catalogue of test particles at $r_\gal$ between 13--14~kpc, for each of the two lopsided warp models: the Sine Lopsided warp (\emph{left}) and the S Lopsided warp (\emph{right}). \emph{Top}: no errors and no Selection Function, \emph{Middle:}  no errors and GaiaDR2 selection function ($G<20$), \emph{Bottom:} GaiaDR2 errors and selection function ($G<20$).}
\label{fig:PCMs_particles}
\end{center}
\end{figure}

The \ngc~PCM signature for the first test scenario, a circular orbit, is shown in Fig.~\ref{fig:PCMs_circular} for each of the three warp models. The Simple Symmetric model produces a single peak (black dot) at a co-latitude corresponding to the warp's tilt. Lopsided models produce double-symmetric contours, since each side of the warp has a different amplitude $\psi_{up/down}$. For the Sine Lopsided model the PCM signature (red line) is a double or bow-tie contour with extremes at co-latitudes corresponding to the up/down warp amplitudes $\psi_{up/down}$. For the S Lopsided model, there is also a double contour (blue line) but, in this case, each peak extends from the north pole to the co-latitude corresponding to the maximum tilt angle of the up/down warp amplitudes. This is clearly illustrated by how the circles of the S Lopsided signature  (blue) touch the extremes of the Sine Lopsided signature bow-tie (red). Notice also that the signature goes back to the PCM center in between the two loops. This is because the warp disappears next to the line-of-nodes for this model.  

The PCM signature for the second test scenario, a random realisation of particles with $r_\gal$ in the range 13--14~kpc, is shown in Fig.~\ref{fig:PCMs_ensemble}. The figure shows that the signatures for the different warp models have similar overall features as in Fig.~\ref{fig:PCMs_circular}, with some differences for the S Lopsided model. 
The SS model produces a single well-defined peak; the SineLop model produces a bow-tie signature caused by two overlapping peaks, barely resolved in this plot; and the SLop model produces two overlapping peaks, in this case with a triangular shape extended in azimuth.  For the first two models, the SS and SineLop, the main PCM peaks stay along the $\phi=270\degr$ meridian, which corresponds to the pole's azimuth of the (straight) line of nodes assumed in this example. On the contrary, for the SLop model, even though the line of nodes is straight, there is signal at azimuths approximately $\pm30\degr$ away from the $\phi=270\degr$ azimuth of the line-of-nodes. This means that, for the SLop model, the azimuth of PCM regions were there is signal does not give the twist angle straight away; instead, the twist angle is given by the azimuth of the meridian along which the signature shows reflection symmetry.  

The PCM signature for the last test scenario, the test particle simulations, is shown in Fig.~\ref{fig:PCMs_particles}. In this figure we show the change in PCM signature going, from top to bottom, from the error-free test particles to the mock catalogue described in Sect.~\ref{s:test_scenarios} with GaiaDR2 simulated errors and selection function. For the SineLop model, the left column of the figure shows the bow-tie signature expected from the previous tests, which is only mildly distorted in the last row due to the effect of the observational errors. For the SLop model the signature differs from that of the ensemble of particles, even in the error-free set (top row). As discussed in Sect.~\ref{sec:methods},  this model's test particle set is in an impulsive regime as the SLop warped potential does not allow the particles to reach statistical equilibrium. Therefore, the PCMs shown here are not to be taken as characteristic of the SLop model itself, but as a guide to what the pole count signature might look like for a population of stars that is not in statistical equilibrium with the SLop model.

\end{appendix}
\label{lastpage}

\end{document}